\documentclass[11pt]{article}%

\usepackage[hyphens]{url}
\usepackage{hyperref}
\usepackage[utf8]{inputenc}
\usepackage[OT2,T1]{fontenc}
\usepackage[french,english]{babel}

\usepackage{amsmath}%
\usepackage{amsfonts}%
\usepackage{amssymb}%
\usepackage{graphicx}
\usepackage{color}
\usepackage{xypic} 
\usepackage{textcomp}
\definecolor{FOB}{rgb}{0.,0.,0.8}
\definecolor{FOR}{rgb}{0.7,0.,0.}

\def\Mapinp{{\color{FOR}{$>$}}}
\def\Cale{\hbox to 0.5cm{\hfill}}
\def\Proc{{\bf proc}}
\def\Local{{\bf local}}
\def\Endproc{{\bf end proc}}
\def\Return{{\bf return}}
\def\Global{{\bf global}}
\def\If{{\bf if}}
\def\Then{{\bf then}}
\def\Else{{\bf else}}
\def\Fi{{\bf end if}}
\def\While{{\bf while}}
\def\Do{{\bf do}}
\def\Enddo{{\bf end do}}
\def\Return{{\bf return}}
\def\HFB{\hfill\break}
\def\Qt{\hbox{\hbox to 0.2ex{\hss}\`{}}\hskip-0.2ex}
\def\Exp{\hbox{\hbox to 0.2ex{\hss}\^{}}\hskip-0.2ex}
\def\Diff{\hbox{\textit{diff}}}
\def\Int{\hbox{\textit{int}}}
\def\With#1{\hbox{\textit{with}(\textit{#1})}}
\def\Hb#1{\hbox{\textit{#1}}}
\def\Qte{\textquotesingle}

\providecommand{\keywords}[1]{\textbf{\textit{Key words. ---}} #1}

\newcounter{thenum}

\newenvironment{theorem}{\medbreak\refstepcounter{thenum}\textsc{Theorem} %
\thethenum. ---  \em  }{\rm }

\newenvironment{Proof}{\smallbreak{\sc Proof.} --- \rm}{\quad\bull\smallskip\rm}

\def\cC{{\cal C}}

\def\cL{{\cal L}}

\def\hO{{\tilde O}}

\def\N{{\bf N}}
\def\R{{\bf R}}

\def\td{{\rm d}}

\def\bull{\vrule height .9ex width .8ex depth -.1ex}

\def\textem #1{{\em #1\/}}

\def\eqref #1{(\ref{#1})}

\DeclareSymbolFont{cyrletters}{OT2}{wncyr}{m}{n}
\DeclareMathSymbol{\Sha}{\mathalpha}{cyrletters}{"58}
\DeclareMathSymbol{\rde}{\mathalpha}{cyrletters}{"64}

\DeclareMathOperator{\arccot}{arccot}

\title{$\hbox to 2pt{\hfill} $ \\[-2cm]
Automatic differentiation of hybrid models\\
Illustrated by Diffedge Graphic Methodology.\\
{\normalsize(Survey)}}

\author{John Masse$^{(\text{a})}$, Clara Masse$^{(\text{b})}$,
  \href{http://www.lix.polytechnique.fr/~ollivier/indexEngl.html}{François
    Ollivier}$^{(\text{c})}$ \\[0.3cm] {\text{\normalsize (a)
      \href{http://www.appedge.com/english/en_accueil.html}{APPEDGE}}} {\text{\normalsize
      18--22, rue d'Arras}}\\ {\text{\normalsize 92000 Nanterre,
      France}} \\[0.3cm] {\text{\normalsize (b)
      Esilv}} {\text{\normalsize 12 Avenue Léonard de
      Vinci}}\\ {\text{\normalsize 92400 courbevoie, France}}
  \\[0.3cm] {\text{\normalsize (c)
      \href{https://www.lix.polytechnique.fr/}{LIX}, UMR CNRS -- École
      polytechnique n$^{{\rm o}}$~7161}}\\ {\text{\normalsize F-91128
      Palaiseau cedex, France}}\\[0.3cm] {\normalsize E-mails:
    john.masse@appedge.com}\\ {\normalsize E-mails:
    clara.masse@devinci.fr}\\ {\normalsize E-mails:
    francois.ollivier@lix.polytechnique.fr}} \date{June 2017}

\begin{document}
\maketitle
\begin{abstract} We investigate the automatic differentiation of
  hybrid models, \textem{viz.} models that may contain delays, logical
  tests and discontinuities or loops. We consider differentiation with
  respect to parameters, initial conditions or the time.  We emphasize
  the case of a small number of derivations and iterated
  differentiations are mostly treated with a foccus on high order
  iterations of the same derivation.  The models we consider may
  involve arithmetic operations, elementary functions, logical tests
  but also more elaborate components such as delays, integrators,
  equations and differential equations solvers. This survey has no
  pretention to exhaustivity but tries to fil a gap in the litterature
  where each kind of of component may be documented, but seldom their
  common use. 

The general approach is illustrated by computer algebra experiments,
stressing the interest of performing differentiation, whenever
possible, on high level objects, before any translation in Fortran or
C code. We include ordinary differential systems with discontinuity,
with a special interest for those comming from discontinuous Lagrangians.

We conclude with an overview of the graphic
methodology developped in the Diffedge software for Simulink hybrid
models. Not all possibilities are covered, but the methodology can be
adapted. The result of automatic differentiation is a new block
diagram and so it can be easily translated to produce real time
embedded programs.
\medskip

We welcome any comments or suggestions of
references that we may have missed.

\end{abstract}


\section*{Introduction} The need to differentiate functions which are
not described by formulas but by computer programs is a classical and
recurrent problem. ``Automatic Differentiation'' (AD) contains two
aspects: on the one hand, one wishes to extend the possibilities of
some existing software, most of the time in order to be able to
optimize the choice of some parameter, on the other hand the question
may already be non trivial when posed before the program is
written. Indeed, computing efficiently some derivatives of functions
that do not admit closed form formulas is a non trivial task. For
example, such functions may be integrals, or solutions of systems of
ordinary differential equations, or be implicitely defined by
algebraic equations. We may moreover compose such functions and in
many pactical situations some logical tests may appear, e.g. when some
physical system must remain in some predefined range of the state
space.

This problem is an old one (see \textit{e.g.} Arbogast
\cite{Arbogast1800}\footnote{Rall \cite{Rall2007} does not hesitate to
  trace it back to  Qin Jiushao's \textit{Mathematical Treatise in
    Nine Sections} (1247).}) that has been widely considered in recent years;
many methods and softwares exist. 
One may refer to \cite{Hoffmann} for an introduction to the
subject. See also Griewank \cite{Griewank2008,GriewankWalther2008}.
The Wikipedia page
\href{https://en.wikipedia.org/wiki/Automatic_differentiation}{Automatic
  differentiation} quotes no less than 17 softwares to differentiate
C/C$^{++}$ programs, 6 softwares for Fortran, 5 for Matlab, 9 for
Python, \dots\ However, these software are limited to functions
defined with elementary functions and for some of them a few matrix
operations. The structure of the language itself can be an other
limitation. What can be done if some high level function, like Matlab
functions ``solve'' or ``dsolve'' are used?  Some years ago, the
software Diffedge was introduced to differentiate functions defined by
bloc diagrams in Matlab/Simulink. This corresponds to the user's need
to stay in the same working environment instead of converting his
model to C, Fortran, \dots\ losing then the possibilities offered by
Simulink, as well as the mathematical semantic of the problem to be
solved.  As a consequence, the mathematical strategies that would have
been used to deal with stiff ODEs, discontinuities are lost that may
lead to poor precision.

It is known, even if automatic differentiation is not as widely used
as it should be, that finite difference methods might lead to poor
results \cite{Elsheikh2012} and in some unpredictable way, as one
cannot know what ``small variation'' should be chosen and, worse, it
is not obvious to tell in case of trouble if the choice made was too
big or too small. And this is especially true for embedded code
dealing with noisy data.  Computing iterated derivatives in that way
is a desperate task.

We would like here to emphasize that using automatic differentiation
on some ``low level'' Fortran or C code that contains subroutines for
solving, say, algebraic systems using a Newton method, or integrating
differential equations, may lead to the same kind of difficulties, so
that one should try, whenever possible, to apply automatic
differentiation at the highest level, e.g. \texttt{solve} or
\texttt{dsolve} functions 
as we may encounter them in computer algebra systems like Maple or as
Simulink blocks.

One wishes to keep as much control as possible on the precision of the
results. Of course, this is difficult, as no one knows in general what
is the actual precision of the computations one performs with a
numerical software. A least, one would like the computation of the
derivative to have a precision $\eta$ comparable to the precision
$\epsilon$ of the function itself, \textit{viz.}
$\eta=O(\epsilon)$. To this regard, brute force conversion to a
low-level programming langage might lead, as we will see, to erroneous
results.

The plan of this paper is first to give a general theoretical methodology
for such situations, \textem{viz.} differentiating implicit functions,
ODE solutions, possibly with delays and discontinuities introduced by
logical tests. Then, we will illustrate it with some Diffedge
examples. The last part is devoted to a few recipes to improve the
accurracy of the result and test the numerical precision of a
software.

We assume that the differentiations are performed with respect to a
limited number of parameters (say in practice $3$ or $5$), meaning
that the choice of reverse accumulation is a legitimate strategy. We
will also consider strategies for computing higher order derivatives
(without any other restriction than time and space complexity). The
question of computing high order time derivatives will also be
considered.

We use here Maple just as an illustration of general
recipes and do not consider the many difficulties for producing
effective implementations in this setting, allowing to get
an output in Fortran, C, Matlab or as a Simulink block diagram.

But we conclude with a description of  the Diffedge Package, designed
by the first author, that shows how automatic differentiation can be
performed inside the Simulink environment, and allowing to deal with
some combinations of the situations described here.

The source code used for all the \href{http://www.lix.polytechnique.fr/~ollivier/AD_Web/Automatic%20differentiation.html}{examples} is available online.

\section{Classical tools}

\subsection{Forward and reverse accumulation}

It is well known that two basic opposite approaches may be used for
automatic derivation of formulas given by programs. Such a program may
be represented by a graph, where all nodes correspond to elementary
functions. The \textit{forward approach} will be best to compute the derivative
of many outputs with respect to one input. In the general
case, let us assume that we have a function of $n$ variable, $x_{1}$,
\dots, $x_{n}$ defined in the following way.\hfill\break
$a_{1}:=f_{1}(x)$\hfill\break
$a_{2}:=f_{2}(x,a_{1})$\hfill\break
(\dots)\hfill\break
$a_{s}:=f_{s}(x,a_{1}, \ldots, a_{s-1})$\hfill\break
where the functions $f_{i}$ are elementary function that will depend
in practice of just one or two of their possible arguments:
e.g. $a_{3}:=x_{2}\times a_{1}$. One will complete this program with
the expressions providing the values 
$$
a_{i,j}:=\frac{\partial f_{i}}{\partial x_{j}}
+\sum_{k=1}^{i-1}\frac{\partial f_{i}}{\partial
  a_{k}}a_{k,j}=\frac{\td f_{i}}{\td x_{j}}. 
$$
One sees that, assuming all functions $f_{i}$ to be $+$, $-$ or $\times$,
one needs at most $4s$ elementary operations to compute the outputs of
the initial $s$ operations program together with their derivatives;
one will go to $5s$ if one uses also division $\div$.

This simple idea, that may be adapted to partial derivatives, was
experimented and described by Wengert in 1964 \cite{Wengert1964}.

The \textit{reverse approach} will be best to compute all partial derivatives
of a single output, that we may assume to be $a_{s}=F(x)$ in the above
program. New values $a_{j}'$ are then computed in reverse
order.\hfill\break
$a_{s-1}':=\partial f_{s}/\partial a_{s-1}$\hfill\break
$a_{s-2}':=a_{s-1}'\times\partial f_{s-1}/\partial a_{s-2}+\partial f_{s}/\partial
a_{s-2}$\hfill\break 
(\dots)\hfill\break 
$a_{j}':=\sum_{k=j+1}^{s-1} a_{k}'\times\partial f_{k}/\partial
a_{j}+\partial f_{s}/\partial a_{j}$
(\dots)\hfill\break 
$x_{j}':=\sum_{k=1}^{s-1} a_{k}'\times\partial f_{k}/\partial
x_{j}+\partial f_{s}/\partial x_{j}$\hfill\break 
The requested number of operations if the $f_{i}$ are elementary
operations $+$, $-$, $\times$ will be again $4s$ to compute all the
partial derivative, and $5s$ including division. Iterations of this
process to get higher order derivatives are possible but inefficient,
leading to an exponential complexity. See also Gower and Gower
\cite{Gower2016} on this issue.

The idea goes back to the pionneering work of Baur and Strassen
\cite{BaurStrassen1985}. See also Morgenstern \cite{Morgenstern1985}.

Finding the best solution to compute the jacobian matrix of a
vector function defined by a given straight-line program is known to
be a difficult problem. 
In particular, if some algebraic relations may exist between partial
derivations, the computation of the Jacobian using a minimal number of
multiplication or addition has been shown by Naumann to be NP-Complete
\cite{Naumann2008}. 

The main idea is to consider the function
$\phi:\R^{n}\mapsto\R^{n}$ defined by
$\phi(x)[i]:=x_{i}\prod_{a\in A_{i}}a$, where the $A_{i}$ are subsets
of some finite set $A$. Computing the diagonal Jacobian matrix amounts
to computing the products $\prod_{a\in A_{i}}a$. It may be shown that
deciding whether this may be done in $s$ operations is equivalent to
testing that the ``ensemble computation'' problem can be solved in $s$
steps: $B_{i}:={a}$, $a\in A$ or $B_{i}:=B_{j_{i}}\cup B_{k_{i}}$ with
$j_{i}, k_{i}<i$ and $B_{j_{i}}\cap B_{k_{i}}=\emptyset$.

On sees that computing the Jacobian with a minimal number of
elementary operation is here equivalent to computing the function
$\phi$ itself in the best way. Our concern here is rather to look for
acceptable ways of computing derivation, starting from a given
implementation of a function, assumed to be reasonable if not optimal.

Practical issues are well illustrated by the special case of the
Hessian. A method relying on a first application of the forward
method to compute the gradient, followed by the reverse one, allows to
obtain the Hessian with a complexity proportional to that of the
initial function \cite{Christianson1992}. The ``edge pushing''
algorithm does this but also avoid useless computation, taking into
accound the symmetry of the Hessian matrix \cite{Gower2012}. Graph
colouring algorithms may be used to take advantage of
sparsity \cite{Gebremedhin05}.

Methods have been proposed to unify forward and reverse methods in
order to look for a good compromize according to the situation. See
\textit{e.g.} \textsc{Volin1985}. 

We will avoid here such questions\footnote{But these are also important
  issues. E.g. in some complicated situations, finite difference
  methods may be unable to decide that some output does not depend on
  a parameter.} and will foccuss on the cases where
some nodes correspond in fact to a complex function, that solves
algebraic equations, integrate differential equations etc. Our choice
will be the direct approach, assuming that in practice one will try to
optimize the behaviour of some system with respect to a limited number
of parameters. But we need first some more brief recalls of classical
methods.

\subsection{An illustration in Maple}

The Maple package \href{http://www.lix.polytechnique.fr/~ollivier/AD_Web/CODEGEN/codegen.html}{Codegen} may be used to produce C code, provided that
the Maple procedure is limited to elementary structures and
functions. It also provides gradient computation using direct and
reverse accumulation (reverse is the default), as well as Jacobian and
Hessian, taking for argument any procedure that returns a number or a
vector of numbers. The \texttt{optimize} function uses Maple's
remember table to identify common subexpression, avoiding to recompute
them. Here is a simple example, showing first the reverse
mode.
\medskip

{\small\parskip=0pt

\noindent\Mapinp\ \With{codegen}\HFB
\Mapinp\ $F:=\Proc(x,y) \Local\> a, b, c, df; a := x+y; b := a*x; c :=
b+2; y*c\>\Endproc:$\HFB
\Mapinp\ $H := GRADIENT(F, mode = reverse)$\HFB
{\color{FOB} $H := \Proc(x, y)\> \Local\> a, b, c, df; a := x+y; b :=
  a*x; c := b+2;$\HFB
\Cale $df := array(1 .. 3); df[3] := y; df[2] := df[3]; df[1] :=
df[2]*x;$\HFB
\Cale $\Return\> a*df[2]+df[1], c+df[1]\> \Endproc$}\Cale $(1)$\HFB
\Mapinp\ $H2 := optimize(H);$\HFB
{\color{FOB} $H2 := \Proc (x, y) \Local\> a, df, t2; a := x+y; df := array(1 .. 3); df[1] := y*x; t2 := df[1];$\HFB
\Cale $\Return a*y+t2, a*x+t2+2\> \Endproc$} \Cale $(2)$\HFB
\Cale\HFB
\Mapinp{\color{FOR} $\sharp$ \textit{We see now the same example with the direct mode.}}\HFB
\Cale\HFB
\Mapinp\ $K := GRADIENT(F, mode = forward)$\HFB
{\color{FOB} $K := \Proc(x, y)\> \Local\> a, b, c, da, db, dc;$\HFB 
\Cale $da := array(1 .. 2); db := array(1 .. 2); dc := array(1
.. 2);$\HFB 
\Cale $da[1] := 1; da[2] := 1; a := x+y; db[1] := x*da[1]+a;$\HFB 
\Cale $db[2] := da[2]*x; b := a*x; dc[1] := db[1]; dc[2] := db[2]; c
:= b+2;$\HFB
\Cale $\Return\> dc[1]*y, y*dc[2]+c$\HFB 
\Endproc}\Cale $(3)$\HFB
\Mapinp\ $K2 := optimize(K)$\HFB
{\color{FOB} $K2 := \Proc(x, y)\>\Local\> a, db, dc;$\HFB
\Cale$db := array(1 .. 2); dc := array(1 .. 2);$\HFB 
\Cale$a := x+y; db[1] := a+x; dc[1] := db[1];$\HFB 
\Cale$\Return dc[1]*y, a*x+y*x+2$\HFB 
\Endproc}\Cale $(4)$
}

More details about this implementation may be found in Monagan and
Neuenschwander's paper \cite{Monagan1993}.

\subsection{Using dual numbers and series}

It is also known that computing the derivative with respect to
$x_{i}$ using the direct approach is equivalent to computing with
``dual numbers'', that is first order truncated power series $a_{0} +
a_{1}x_{i}+ O(x_{i}^{2})$. This idea allows to compute iterated
partial derivatives of arbitrary order with respect to different
inputs with quite a good complexity, using power series truncated at a
higher order $\sum_{j=0}^{r} a_{j} x_{i}^{j} + O(x_{i}^{r+1})$. 
Up to logarithmic terms, the
multiplication of power series has a linear complexity in the size of
the data\footnote{It is said to be \textit{soft
    linear}. \textit{E.g.}, we will write $\ln\,n\left(\ln\,ln\,
  n\right)n=\hO(n)$.}, \textit{i.e.} the number of monomials, using
dense
representation and to some extend also sparse representation
\cite{HoevenLecerf2013,LebretonSchost2016}.

\begin{theorem} Given a straight line program of size $L$ that
  computes a function $F(x_{1}, \ldots, x_{s})$ we may compute all
  derivatives 
$$
\frac{\partial^{\sum_{i=1}^{s}\ell_{i}} F}{\prod_{i=1}^{s}\partial x_{i}^{\ell_{i}}},\quad
  \ell_{i}\le \alpha_{i} 
$$
in $\hO(\prod_{i=1}^{s}\alpha_{i})L$.
\end{theorem}

One may remark that the iterated use of the reverse approach leads to
an exponential complexity.

One may also generalize this idea to the case of multiple derivations,
using again truncated first order power series: $a_{0}+\sum_{i=1}^{n} a_{i}x_{i}
(x_{i}|1\le i\le n)^{2}$. But computing higher order derivatives using
series truncated at order $r+1$: $\sum_{|\alpha|\le r}
a_{\alpha}x^{\alpha}(x_{i}|1\le i\le n)^{r+1}$, one may not escape an
exponential growth of the ouput.  Such a structure is just equivalent to
jets of order $r$.

In any typed computer algebra system, jet space may be defined for any
ring and provide the basic setting for forward automatic
differentiation of any expression including the basic ring or field
operation and some elementary functions. This has been done in Axiom
by Smith \textit{et al.} \cite{Smith2007}.

\section{Our theoretical setting illustrated using computer algebra}

In this section, we explain how to extend automatic differentiation to
more complicated situations and, before comming to Simulink models, we
illustrate our ideas with computer algebra tools. One will soon see
that it is more complicated that just computing closed formulas and
then differentiating them, as very often closed formulas do not exist.

\subsection{Introducing new functions}

One often needs to introduce new functions, satisfying some ordinary
or partial differential systems. E.g. some functions $F_{i}$ are such
that 
$$
\frac{\partial F_{i}}{\partial x_{j}} = G_{i,j}(F,x).
$$
From a mathematical standpoint, the solution is straightforward. From
the standpoint of practical implementation, the problem may be less
obvious if one does not want to reimplement completely the
differentiation. Using typed systems like Axiom, it is easy to design
a new package with the requested functions and their differentiation
rules. For example, here is a part of the package
\texttt{ElementaryFunction} written by Manuel Bronstein\cite{Bronstein1987}.
\medskip

{\tiny
\begin{tabular}{|c|}\hline
\vbox{\hsize=12.5cm
\begin{verbatim} 
    oplog := operator("log"::Symbol)$CommonOperators
    opexp := operator("exp"::Symbol)$CommonOperators
    opsin := operator("sin"::Symbol)$CommonOperators
    opcos := operator("cos"::Symbol)$CommonOperators
    optan := operator("tan"::Symbol)$CommonOperators
\end{verbatim}
}\\ 
(\dots)\\
\vbox{\hsize=12.5cm
\begin{verbatim} 
    derivative(opexp, exp)
    derivative(oplog, inv)
    derivative(opsin, cos)
    derivative(opcos, - sin #1)
    derivative(optan, 1 + tan(#1)**2)
\end{verbatim}
}\\ \hline
\end{tabular}}
\medskip

Each object in Axiom as a name and a type, and the functions to be
used for them depend on their type too, specified with a
\texttt{\$}. We see here how the derivatives of the operators
\texttt{log}, \texttt{exp}, \texttt{sin} etc.\ that belong to the
category \texttt{CommonOperators} can be defined.

Following such examples, one is free to design functions of one's own
in a new package, containing the requested rules for
derivation. Anyway, there is a price to pay for that liberty, as
programming in such a system may become tedious when one needs to
specify the type of each mathematical object and to specify also how
to change the type of some object according to our goals. An advantage
for this extra work is that one must adopt a cautious programmation
style that avoids many mistakes.
\bigskip

We mention this possibility for the sake of completeness and the
satisfaction of competent and courageous readers.  In Maple, one just
needs to know that the internal \href{http://www.lix.polytechnique.fr/~ollivier/AD_Web/DIFF/Diff.html}{\texttt{diff}} function of Maple, when
encoutering a function \texttt{F} first looks if \texttt{diff/F} is
defined, so that one just has to define it with the proper value.
Then, this new definition will be used by the system to compute the
derivatives of all formulas.

The rule is the following: assuming that \texttt{F} takes $n$
arguments, \texttt{diff/F} is
a procedure, depending of $n+1$ arguments $a_{1}$, \dots, $a_{n}$,
$b$, that provides the value of $\partial F(a_{1}, \ldots, a_{n})/\partial b$.

E.g, assume we want to define two new functions $f(x,y)$ and
$g(x,y)$ such that 
$$
\frac{\partial f}{\partial x}= 2g, \frac{\partial f}{\partial y}= 3g,
\frac{\partial g}{\partial x}= -2f, \frac{\partial g}{\partial y}= -3f.
$$
(Of course $\sin(2x+3y+c_{1})$ and $\cos(2x+3y+c_{2})$ are
solutions\dots\ it is just to consider a simple example.)
A possible answer is illustrated by the following Maple session.

\medskip

{\small\parskip=0pt

\noindent\Mapinp\ $\Qt \Diff/f\Qt:=\Proc(a,b,c)\>
\Diff(a,c)\ast2\ast g(a,b)+\Diff(b,c)\ast3\ast g(a,b)\>\Endproc:$\HFB
\Mapinp\ $\Qt \Diff/g\Qt:=\Proc(a,b,c)\>
-\Diff(a,c)\ast2\ast f(a,b)-\Diff(b,c)\ast3\ast g(a,b)\>\Endproc:$\HFB
\Mapinp\ $\Diff(f(a(c),b(c)),c)$\HFB
\Cale\Cale{\color{FOB} $2\left(\frac{\td}{\td c}a(c)\right)g(a(c),b(c))+
\left(\frac{\td}{\td c}b(c)\right)g(a(c),b(c))3$}\HFB
\Mapinp\ $\Diff(f(x,y)\ast g(x,y),x)$\HFB
\Cale\Cale{\color{FOB}$2g(x,y)^{2}-2f(x,y)^{2}$}\HFB
\Mapinp\ $\Diff(\Int(f(x,y), x=a..b),y)$\HFB
\Cale\Cale{\color{FOB}$\int_{a}^{b}3g(x,y)\td x$}\HFB
\Mapinp\ $\Diff(f(z,z\ast\ast2),z)$\HFB
\Cale\Cale{\color{FOB}$2g(z,z^{2})+6zg(z,z^{2})$}\HFB
}

Before using such a possibility in a computer algebra system, one must
be sure of what we are doing: partial derivatives here are assumed to commute.
An other way to proceed in Maple would be to use the Diffalg or the
\href{http://www.lix.polytechnique.fr/~ollivier/AD_Web/DIFFALG/Diffalg.html}{DifferentialAlgebra} packages and compute
first a \textit{``characteristic set of the prime differential ideal generated
by the above system''}. Basically it is some kind of normalized set of
equations, which may be difficult to compute but provides any information
one may need. 

In our case, the result will be the system itself, because the partial
derivations $\partial/\partial x$ and $\partial/\partial y$ do commute
and the computation is reduced to checking this.
$$
\frac{\frac{\partial f}{\partial x}}{\partial y}
=\frac{\frac{\partial f}{\partial y}}{\partial x}\quad\hbox{and}\quad
\frac{\frac{\partial g}{\partial x}}{\partial y}
=\frac{\frac{\partial g}{\partial y}}{\partial x}.
$$
We can then use \texttt{NormalForm} to replace the derivatives with
their proper values.

\medskip

{\small\parskip=0pt

\noindent\Mapinp\ $\With{DifferentialAlgebra}:\> \With{Tools}:$\HFB
\Mapinp\ $R:=\Hb{DifferentialRing}(\Hb{Blocks}=\Hb{lex}[f,g],
\Hb{derivations}=[x,y]):$\HFB
\Mapinp\ $\Hb{syst}:=[\Hb{Differentiate}(f,x,R)=2\cdot g,
\Hb{Differentiate}(f,y,R)=3\cdot g,$\HFB
\phantom{\Mapinp\ $\Hb{syst}:=[$}$\Hb{Differentiate}(g,x,R)=-2\cdot f,
\Hb{Differentiate}(g,y,R)=-3\cdot f]:$\HFB
\Mapinp\ $\Hb{simplified\_syst}:=\Hb{RosenfeldGroebner}(\Hb{syst},R):$\HFB
\Mapinp\ $\Hb{Equations}(\Hb{simplified\_syst})$\HFB
\Cale\Cale\Cale{\color{FOB} $[[f_{x}-2g,g_{x}+2f,f_{y}-3g,g_{y}+3f]]$}\HFB
\Mapinp\ $\Hb{NormalForm}(\Hb{Differentiate}(f\cdot
g,x,R),\Hb{simplified\_syst})$\HFB
\Cale\Cale\Cale{\color{FOB} $[-2f^{2}+2g^{2}]$}\HFB
\Mapinp\ $\Hb{NormalForm}(f,\Hb{simplified\_syst})$\HFB
\Cale\Cale\Cale{\color{FOB} $f$}
}
\medskip

This might look complicated but it is good to know that such tools
exist in case of need, provided that one keeps in mind that they could be
time and memory consuming. They must be
reserved to cases were one does not know \textit{a priori} a normal
form for our relations. Moreover, we need to use here the
\texttt{Differentiate} function of the \texttt{DifferentialAlgebra}
package, which is different from the \texttt{Diff} function used
elsewhere. 

Diffalg may also be used to \textit{eliminate} some function, using
specific orderings on derivatives. In the next example
\texttt{blocks=[lex[f],lex[g]]} means that $f$ and all its derivatives
are greater than $g$ and all its derivatives. So, one does not express
$\partial g/\partial y$ as $-3f$ any more, but instead $f$ is
expressed as $(-1/3)(\partial g(x,y)/\partial y$.

\medskip

{\small\parskip=0pt

\noindent\Mapinp\ $\With{DifferentialAlgebra}:\> \With{Tools}:$\HFB
\Mapinp\ $R:=\Hb{DifferentialRing}(\Hb{Blocks}=\Hb{lex}[f], \Hb{lex}[g],
\Hb{derivations}=[x,y]):$\HFB
\Mapinp\ $\Hb{syst}:=[\Hb{Differentiate}(f,x,R)=2\cdot g,
\Hb{Differentiate}(f,y,R)=3\cdot g,$\HFB
\phantom{\Mapinp\ $\Hb{syst}:=[$}$\Hb{Differentiate}(g,x,R)=-2\cdot f,
\Hb{Differentiate}(g,y,R)=-3\cdot f]:$\HFB
\Mapinp\ $\Hb{simplified\_syst}:=\Hb{RosenfeldGroebner}(\Hb{syst},R):$\HFB
\Mapinp\ $\Hb{Equations}(\Hb{simplified\_syst})$\HFB
\Cale\Cale\Cale{\color{FOB} $[[g_{y}+3f,3g_{x}-2g_{y},g_{y,y}+9g]]$}\HFB
\Mapinp\ $\Hb{NormalForm}(\Hb{Differentiate}(f\cdot
g,x,R),\Hb{simplified\_syst})$\HFB
\Cale\Cale\Cale{\color{FOB} $[2g^{2}-\frac{2}{9}g_{y}^{2}]$}\HFB
\Mapinp\ $\Hb{NormalForm}(f,\Hb{simplified\_syst})$\HFB
\Cale\Cale\Cale{\color{FOB} $-\frac{1}{3}g_{y}$}
}
\medskip

Longer considerations would exceed the ambitions of this paper, but
these tools deserve to be mentioned
here. For more details, one may refer to the work of Boulier, the
first designer of \texttt{Diffalg} \cite{BLOP2,Boulier2001} and of
Hubert who developped the latest Maple versions
\cite{Hubert2003,Diffalg} and considered also \textit{non commuting
  derivations} \cite{Hubert2005}. The new Maple package
\texttt{Differentialalgebra} is based on Boulier's
\textit{Bibliothèques Lilloises d'Algèbre Différentielle} cite{BLAD},
redesigned for Maple by Boulier and Cheb-Terrab.
\bigskip

Using numerical schemes obtained in this way, it is easily seen that, 
knowing an approximation of some
function $f(x_{1}, \ldots, x_{n})$ with an error bounded by
$\epsilon$, the approximation for any iterated derivative of $f$ will
be $O(\epsilon)$. However, one may expect an exponential growth with
respect to the order of derivation. The point for us is that we are
sure that increasing the precision on the computation of the $f_{i}$
will increase the precision on the computation of derivatives, which
may not be granted with other tools.
\bigskip

\subsection{Runge--Kutta methods}

Assume indeed that one computes some function $f$ that is the solution
of some ordinary differential equation: $f'=G(f)$. We may compute an
approximation of $f$ for $x\ge 0$ using a \href{http://www.lix.polytechnique.fr/~ollivier/AD_Web/RUNGEKUTTA/RungeKutta.html}{Runge-Kutta} method of order
$n$, that is implemented in some low-level C or Fortran Program, which
would be a common choice. If $G$ is linear (resp.\ quadratic), $f$
will be approximated by a piecewise polynomial function of degree $n$
(resp.\ $2^{n}-1$). So, we see that we have no hope to evaluate
derivatives of order higher that this degree from the implemented
Runge-Kutta approximation, even if we increase its precision with a
smaller step. We need at least to change the order of the Runge Kutta
method itself! And \textit{A priori}, we can only expect acceptable
values for derivatives up to order $n$, at most: all greater
derivatives will be $0$ with a linear equation and the approximations
obtained for non linear equations are not very good.

As an example, consider $f(x):=(1+x)^{-1}$, solution
of $f'=G(f):=-f^{2}$ with $f(0)=1$. The following table gives the
successive devives of $f(x)$ and their approximation using
``midpoint'' and RK4 methods.
\smallskip

{\tiny
\noindent\begin{tabular}{lrrrrrrrrrrr}
Order of derivation $i$&1  & 2&    3&  4&      5&   6&       7&     8&         9&10&11\\ 
$f^{(i)}(0)$&           $-1$& 2& $-6$& 24& $-120$& 720& $-5040$& 40320& $-3.63 10^{5}$& $3.63 10^{6}$& $-3.99 10^{7}$\\
Midpoint&              $-1$& 2& $-1.5$&0&0      &0   &0       &0     &0     &       0&    0\\
RK4&                   $-1$& 2&$-6$&$24$& $-115$& 600& $-3438.75$&19530&$-1.0773
10^{5}$&$5.481 10^{5}$&$-2.44 10^{6}$\\
\end{tabular}
}
\smallskip

Even if the use of AD for computing
derivatives of order greater than $2$ is uncommon, it seems worth to
mention this 
limitation. We will return to the derivation of solutions of
differential equation in section~\ref{Diff-eq}

\subsection{Conditionals and piecewise functions}

As noticed by Beck and Fischer \cite{Beck1994}, the differentiation of
piecewise functions may lead to erroneous results in some cases. To
summarize the situation, assume that a real function is defined on the
union of disjoints intervals $\bigcup_{i=1}^{s} I_{i}$, where
$I_{i}=]/[a_{i}, b_{i}]/[$ are intervals with a nonempty interior,
    that is $a_{i}<b_{i}$. We may also admit  $\pm\infty$ values for
    the intervals bounds. In such a case, assuming that a function $F$
    is described on $I_{i}$ by some program that implements a
    differentiable function
    $f_{i}$ and that this program can be differentiated, the only
    trouble can arise if $a_{i}=b_{i+1}$: then the value of the
    derivatives must agree. If not, the function is not differentiable
    at $a_{i}$

The real difficulty may be illustrated by the \href{http://www.lix.polytechnique.fr/~ollivier/AD_Web/PIECEWISE/piecewise.html}{following example}:
$F(x):= (1-\cos(x))/x$ if $x\neq0$, $F(0)=0$. Most implementations will
provide $F'(x)=\sin(x)/x-(1-\cos(x))/x^{2}$ if $x\neq0$, $F'(0)=0$: even if
the function admits a continuous derivative, the value computed for
$x=0$ is erroneous. Let us look at Maple possibilities. 

A naïve use of the \texttt{diff} function gives poor results. 
In principle \texttt{D(F)(x)}
and \texttt{diff(F(x),x)} should be equivalent: their internal
representation is different, but \texttt{D(F)(x) - diff(F(x),x)}
simplifies to $0$. Nevertheless, \texttt{Diff} and \texttt{D} act in
different ways. Using \texttt{diff(F(x),x)}, \texttt{F(x)} is
evaluated first, and then the differentiation performed\dots\ It is
best here to use the \texttt{D} function, more adapted to the
differentiation of a procedure. 
\smallskip

{\small\parskip=0pt

\noindent\Mapinp\ $F:=\Proc(x)$ \If\ $x\neq0$
\Then\ $\frac{1-\cos(x)}{x}$ \Else\ $0$ \Fi\ \Endproc:\HFB
\Mapinp\ $\Diff(F(x),x);$\HFB
\Cale\Cale\Cale\Cale$\frac{\sin(x)}{x}-\frac{1-\cos(x)}{x^{2}}$\HFB
\Mapinp${\rm D}(F)$\HFB
$\Proc(x)$\HFB
\Cale$\If\> x<>0\>\Then\> \sin(x)/x - (1-\cos(x))/x\Exp2\>\Else\>0\>\Fi$\HFB
\Endproc
}
\smallskip

One sees that the result at $x=0$ is wrong, as suspected. However, one
may easily change the definition of the function, using the full
possibilities of computer algebra, to be sure it can be
differentiated up to a chosen order, here $10$.
\smallskip

{\small\parskip=0pt

\noindent\Mapinp\ $F:=\Hb{subs}\left(T=\Hb{simplify}\left(
\frac{\textstyle\Hb{convert}(\Hb{taylor}(1-\cos(x),x=0,12),polynom)}
{\textstyle  x}\right)\right.$,\HFB
\hphantom{\Mapinp\ $F:=\Hb{subs}\left(\right.$} 
$\Proc(x)$ \If\ $x\neq0$ \Then\ $\frac{1-\cos(x)}{x}$ 
\Else\ $0$ \Fi\ \Endproc$\vphantom{\frac{1}{x}}\left.\right)$;\HFB
{\color{FOB}$F:=\Proc(x)$\HFB
\Cale\If\ $x<0$ \Then\HFB
\Cale\Cale$(1-\cos(x))/x$\HFB
\Cale\Else\HFB
\Cale\Cale$1/3628800\ast x\ast(x\Exp8 -90\ast x\Exp6+5040\ast x\Exp4
-151200\ast x\Exp2+1814400)$\HFB
\Cale\Fi\HFB
\Endproc\HFB}
\Mapinp${\rm D}(F)$\HFB
{\color{FOB}$\Proc(x)$\HFB
\Cale$\If\> x<>0$ \Then\HFB
\Cale\Cale$\sin(x)/x - (1-\cos(x))/x\Exp2$\HFB
\Cale\Else\HFB
\Cale\Cale$1/3628800\ast x\Exp8 -1/40320\ast x\Exp6+1/720\ast x\Exp4
-1/24\ast x\Exp2+1/2$\HFB
\Cale\Cale$1/3628800\ast(8\ast x\Exp7 -540\ast x\Exp5+20160\ast x\Exp3
-302400\ast x)$\HFB
\Cale\Fi\HFB
\Endproc}
}
\smallskip

Then, we can use the facilities of the \texttt{CodeGeneration} package
to translate this Maple function in C or Fortran.
\smallskip

{\small\parskip=0pt

\noindent\Mapinp\\textit{With}(\textit{Codegeneration}): \HFB
\noindent\Mapinp\\textit{Fortan}$(F)$;\HFB
\texttt{\color{FOB}
\Cale\Cale\Cale\HFB doubleprecision function F (x)\HFB
\Cale\Cale\Cale\Cale integer x\HFB
\Cale\Cale\Cale\Cale if (x .ne. 0) then\HFB
\Cale\Cale\Cale\Cale\Cale F = (1 - cos(dble(x))) / dble(x)\HFB
\Cale\Cale\Cale\Cale\Cale return\HFB
\Cale\Cale\Cale\Cale else\HFB
\Cale\Cale\Cale\Cale\Cale F = dble(x * (x ** 8 - 90 * x ** 6 + 5040 * x
** 4\HFB
\Cale\Cale\Cale\Cale\Cale\Cale - 151200 * x ** 2 + 1814400))/0.362880D7\HFB
\Cale\Cale\Cale\Cale\Cale return\HFB
\Cale\Cale\Cale\Cale end if\HFB
\Cale\Cale\Cale end\HFB
}}
\smallskip

Maple can also handle more easily such situations using \texttt{piecewise}.

\smallskip

{\small\parskip=0pt
\noindent\Mapinp
$G(x):=piecewise\left(x\neq0,\frac{1-\cos(x)}{x},0\right):$ \HFB
\noindent\Mapinp\ $\Diff(G(x),x);$ \HFB
\texttt{\color{FOB}
\Cale\Cale\Cale\Cale\Cale\Cale$\left\{\begin{array}{cc}
\frac{1}{2} & x=0\\
\frac{\sin(x)}{x} - \frac{1 -\cos(x)}{x^{2}}& otherwise
\end{array}\right.$\HFB
}
\noindent\Mapinp\ $D[1](G)$;\HFB
\texttt{\color{FOB}
\Cale\Cale\Cale\Cale\Cale\Cale$x\rightarrow piecewise\left(x=0,\frac{1}{2},\frac{\sin(x)}{x} - \frac{1 -\cos(x)}{x^{2}}\right)$\HFB
}
\noindent\Mapinp\ $D[1,1,1](G)$;\HFB
\texttt{\color{FOB}
\Cale\Cale\Cale\Cale\Cale\Cale$x\rightarrow piecewise\left(x=0,\frac{1}{4},-\frac{\sin(x)}{x} -\frac{3\cos(x)}{x^{2}}+\frac{6\sin(x)}{x^{3}}-\frac{6(1-\cos(x))}{x^{4}}\right)$
}
}

The derivations are computed in the right way, up to any order,
without any intervention of the user. However, the Fortran translation
is not suited for automatic differentiation.

{\small\parskip=0pt

\noindent\Mapinp\\textit{With}(\textit{Codegeneration}): \HFB
\noindent\Mapinp\\textit{Fortan}$(F)$;\HFB
\texttt{\color{FOB}
\Cale\Cale\Cale\HFB doubleprecision function F (x)\HFB
\Cale\Cale\Cale\Cale integer x\HFB
\Cale\Cale\Cale\Cale if (x .ne. 0) then\HFB
\Cale\Cale\Cale\Cale\Cale G = (0.1D1 - cos(dble(x))) / dble(x)\HFB
\Cale\Cale\Cale\Cale\Cale return\HFB
\Cale\Cale\Cale\Cale else\HFB
\Cale\Cale\Cale\Cale\Cale G = 0.0D0\HFB
\Cale\Cale\Cale\Cale\Cale return\HFB
\Cale\Cale\Cale\Cale end if\HFB
\Cale\Cale\Cale end\HFB
}}
\smallskip

 A choice like the
following, although the formula is discontinuous at $x=0.1$, will
allow differentiation and will be also more accurate for small values
of $x$.

{\small\parskip=0pt

\noindent\Mapinp\ $H:=\Hb{subs}\left(T=\Hb{simplify}\left(
\frac{\textstyle\Hb{convert}(\Hb{taylor}(1-\cos(x),x=0,12),polynom)}
{\textstyle  x}\right)\right.$,\HFB
\hphantom{\Mapinp\ $F:=\Hb{subs}\left(\right.$} 
$\Proc(x)$ \If\ $|x|>0.1$ \Then\ $\frac{1-\cos(x)}{x}$ 
\Else\ $T$ \Fi\ \Endproc$\vphantom{\frac{1}{x}}\left.\right)$;\HFB
{\color{FOB}$F:=\Proc(x)$\HFB
\Cale\If\ $x<0$ \Then\HFB
\Cale\Cale$(1-\cos(x))/x$\HFB
\Cale\Else\HFB
\Cale\Cale$1/3628800\ast x\ast(x\Exp8 -90\ast x\Exp6+5040\ast x\Exp4
-151200\ast x\Exp2+1814400)$\HFB
\Cale\Fi\HFB
\Endproc\HFB}
}
\smallskip

{\small\parskip=0pt

\noindent\Mapinp\\textit{Fortan}$(H)$;\HFB
\texttt{\color{FOB}
\Cale\Cale\Cale\HFB doubleprecision function F (x)\HFB
\Cale\Cale\Cale\Cale integer x\HFB
\Cale\Cale\Cale\Cale if (0.1D0 .lt. dble(abs(x))) then\HFB
\Cale\Cale\Cale\Cale\Cale F = (1 - cos(dble(x))) / dble(x)\HFB
\Cale\Cale\Cale\Cale\Cale return\HFB
\Cale\Cale\Cale\Cale else\HFB
\Cale\Cale\Cale\Cale\Cale F = dble(x * (x ** 8 - 90 * x ** 6 + 5040 * x
** 4\HFB
\Cale\Cale\Cale\Cale\Cale\Cale - 151200 * x ** 2 + 1814400))/0.362880D7\HFB
\Cale\Cale\Cale\Cale\Cale return\HFB
\Cale\Cale\Cale\Cale end if\HFB
\Cale\Cale\Cale end\HFB
}}
\smallskip

One sees on this apparently simple problem of AD can be very
difficult when acting on some low level code that has not been
properly implemented in order to facilitate it. The only way then to
avoid inaccurate results would be to reconstruct, if it is still
possible, the original symbolic formula. One may refer to
Shamseddine and Berz \cite{Shamseddine1996} for more details.

\subsubsection{Differentiating flat outputs} 

\textem{Flat systems}\label{Flat}
\cite{fliess,fliess2,fliess3,Sira-Ramirez2004,Levine2009} are examples
of differential control systems the solutions of which may be
parametrized by some functions of their state, called \textem{flat
  outputs} and some derivatives of these functions. A classical
example is that of a
\href{http://www.lix.polytechnique.fr/~ollivier/AD_Web/FLATNESS/Car.html}{car},
described by the following equations:
$$
\begin{array}{ll}
x' &= u \cos \theta \\
y' &= u \sin \theta \\
\theta' &= \frac{u}{\ell} \tan \phi
\end{array}
$$ The functions $x$ and $y$ are easily seen to be flat outputs, as
$\theta=\arctan(y'/x')$ (or $\theta=\pi-\arccot(x'/y')$),
$u=\sqrt{(x')^{2}+(y')^{2}}$ and $\phi=\arctan((\ell/u)\theta')$. So, we
can compute the trajectory of the car knowing $x$ and $y$, provided
that $(x',y')\neq(0,0)$, see \href{http://www.lix.polytechnique.fr/~ollivier/AD_Web/FLATNESS/animation1.gif}{ex.~1}. If the
car follows a trajectory that starts forward and end backward, that is
common for parking, one needs to cross a singularity where both $x'$
and $y'$ vanish, so that $\theta$ is undefined. Such singularities
have been studied in \cite{Kaminski2017}. If $\theta'\neq0$, this is an
\textem{apparent singularity}, as one may use some different flat
outputs, \textit{e.g.} $\theta$ and $\sin(\theta)x-\cos(\theta)y$. See
\href{http://www.lix.polytechnique.fr/~ollivier/AD_Web/FLATNESS/animation2.gif}{ex.~2}. But
this case is mostly theoretical, as in such a situation, one need have
$\phi=\pm\pi/2$. In such a situation, the curve parametrized by
$(x,y)$ is a cusp: $x(t)\simeq At^{2}$ and $y(t)\simeq Bt^{3}$. 

With an actual car, a common limit is $|\phi|>\pi/4$. In such a case,
we have, \textit{e.g.} $x(t)\simeq At^{2}$ and $y(t)\simeq Bt^{5}$,
that is a \textem{rhamphoid cusp}. This corresponds to an essential
singularity, meaning that parametrization cannot be achieved by
choosing some new flat outputs. An efficient pratical solution is then
to compute power series solutions for $\theta$ and $\phi$ at the
singular point, as described in the Maple worksheet. See
\href{http://www.lix.polytechnique.fr/~ollivier/AD_Web/FLATNESS/animation3.gif}{ex~3}. Just
using the formula with rounded floating point numbers near the
singularity, we have no division by $0$ and the value obtained for
$\theta$ is satisfactory. That obtained for $\phi$ is erratic. see \href{http://www.lix.polytechnique.fr/~ollivier/AD_Web/FLATNESS/animation4.gif}{ex.~4}. 

\subsection{Loops. Solving ``algebraic'' equations}

Algebraic\footnote{By ``algebraic'', we mean here ``non differential'';
  such equations may not be polynomial.} equations solvers are a
  second example of high level functions that require a special care to
  be differentiated. In low level code, they will appear as loops,
  such as \texttt{until $|F(x)|<\epsilon$ repeat [\dots]}, whereas in
  Maple one would encounter \texttt{solve(F(x)=0,x)}. However, the
  \texttt{solve} function is not recognized by Maple's \texttt{C}
  function. Applying \texttt{diff} to the result of \texttt{solve},
  which may be a sequence, can be
  disappointing. One need also notice that the numerical resolution of
  a polynomial system may be, in the general case, as hard as its symbolic
  solving. If we consider more 
  general situations, like polynomials in $x$ and $e^{x}$, one can
  in the best cases provide bounds on the number of solutions.  

Anyway, we take here for granted the existence of a reasonable
numerical method for a given system, that is a starting point in a ball
containing a single simple root, from where some iteration process---say a
Newton method for a one variable system---will converge. 

Let $(x_{1}(y), \ldots, x_{n}(y))$ be a regular solution
of a system $P_{i}(x_{1}, \ldots,
x_{n},y)=0$, $1\le i\le n$, that is a $n$-uple of functions such that
$P_{i}(x(y),y)=0$ and $|J_{P}|(x(y),y)\neq0$ on the definition domain
of the $x_{i}$, where $J_{P}:=(\partial
P_{i}/\partial x_{j})$. Whatever can be the numerical method used to
compute the $x_{i}$, one knows that 
\begin{equation}\label{implicit}
\frac{\partial \eta_{j}}{\partial
  \theta}=J_{P}^{-1}(\eta)\left(\begin{array}{c}\frac{\partial
    P_{1}}{\partial \theta}(\eta)\\ \vdots \\ \frac{\partial
    P_{n}}{\partial \theta}(\eta)\end{array}\right).
\end{equation}
 This formula may be regarded as a differential equation, so that we
need not repead what was said above\footnote{Let us just mention, for
  whoever would need to compute high order derivatives, that
  Bostan \textit{et al.} \cite{Bostan2007} have shown that, for a
  single function defined by a polynomial equation $P(x,y)$, one could
  in fact define $x$ as the solution of a \textit{linear} equation,
  allowing to compute iterated derivatives faster that using Newton's
  method for series.}.

From now one, this topic will be illustrated with the case of a single
equation in a single variable. A straightforward implementation of
\href{http://www.lix.polytechnique.fr/~ollivier/AD_Web/NEWTON/Newton2.html}{Newton's method} in Maple may look like the following procedure \texttt{g}.
It is well known that, for computing square roots $\sqrt{a}$, the
Newton method iterates the operator $x\mapsto
(a-x^{2})/2x=(1/2)(a/x+x)$. 

{\small\parskip=0pt\parindent=0pt

\Mapinp\ $g\> :=\> \Proc (a, epsilon, c)\> \Local\> x,\> b$;\HFB
\Cale\Cale \If\  $type(c, numeric)$ \Then\ $x := c$ \Else\ $x := 1.$ \Fi;\HFB 
\Cale\Cale \If\  $type(a+epsilon+c, numeric)$ \Then \HFB
\Cale\Cale\Cale $b := a/x$;\HFB 
\Cale\Cale\Cale \While\  $evalb(epsilon < abs((b-x)/x))$ \Do\HFB 
\Cale\Cale\Cale\Cale  $x := (1/2)*x+(1/2)*b$; $b := a/x$ \HFB
\Cale\Cale\Cale \Enddo;\HFB 
\Cale\Cale\Cale $x$ \HFB
\Cale\Cale \Else\ \Return\ \Qte$procname(args)$\Qte\ \Fi \HFB
\Cale\Cale \Endproc
}
\smallskip

{\small\parskip=0pt\parindent=0pt

\Mapinp\ \Qt$diff/g$\Qt\ $:=$ \Proc$(a, epsilon, c, t) (1/2)*(diff(a,
t))/g(a, epsilon, c)$ \Endproc 
}
\smallskip

{\small\parskip=0pt\parindent=0pt
\Mapinp\ $eval(subs(a = 1.5, diff(g(a, 0.1e-3, 1.00001), a, a))), eval(subs(a =
1.5, diff(a^(1/2), a, a)))$\HFB
\Cale\Cale\Cale\Cale {\color{FOB}-.1360827546, -.1360827636}
}

One sees that, with a proper definition of \texttt{diff/g}, iterated
differentiation works well.  

It is know that the limit of derivatives of the sequence of functions
associated to Newton's method converges to the derivative of the limit
algebraic function (see e.g. Beck \cite{Beck1994b}). However, in
practice difficulties arize as a limited number of iterations will be
achived.  Even in cases where the direct automatic differentiation of
the numerical code can produce accurate values, it is likely to be
slower; see Bell and Burke \cite{Bell2008} and the reference
therin. One must however stress that even if the numerical
approximation is very good, its direct derivation may lead to poor
results, as shown by the following example.

\smallskip

{\small\parskip=0pt\parindent=0pt

\Mapinp\ plot(g(a, 0.5e-1, 1.), a = .1 .. 2.000)
}
\smallskip

\includegraphics[width=4cm]{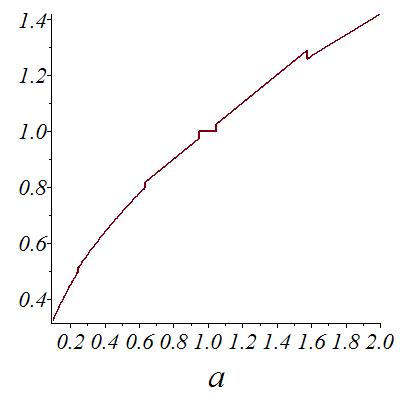}

Arround the initial value, the function is flat, and a single Newton
step produces a linear function, so that the second derivative will be
$0$. Very often, such procedures are used in sequence and one improves
the efficiency by starting with the previously computed value, as in
the following ``memory'' function.

{\small\parskip=0pt\parindent=0pt

\Mapinp\ $g2:=\Proc(a,epsilon)\> \Global\> p\_res;\> \Local\> x, b$;\HFB
\Cale\Cale\Cale\Cale  \If\ $type(p\_res,numeric)$ \Then\ $x:=p\_res$
\Else\  $x:=1$ \Fi;\HFB 
\Cale\Cale\Cale\Cale  \If\ $type(a+epsilon,numeric)$ \Then\HFB  
\Cale\Cale\Cale\Cale\Cale  $b:=a/(x);$\HFB 
\Cale\Cale\Cale\Cale\Cale  \While\ $evalb(abs((b-x)/(x))>epsilon)$
\Do\  $x:=(x+b)/(2)$; $b:=a/(x)$ \Enddo; p\_res:=x\HFB 
\Cale\Cale\Cale\Cale  \Else\ \Return\ \Qte $procname(args)$\Qte\ \Fi\HFB
\Cale\Cale\Cale\Cale  \Endproc
}
\smallskip

Then, the resulting curve for a small value of the precision
$\epsilon$ looks smooth, but it is in fact a piecwise constant
function, so that any attempt to differentiate it will fail. 

\hbox to
\hsize{\hss\includegraphics[width=4cm]{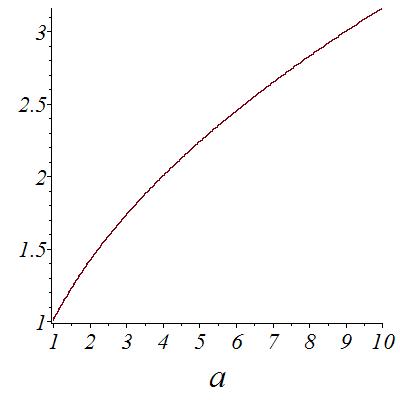}\hss\includegraphics[width=4cm]{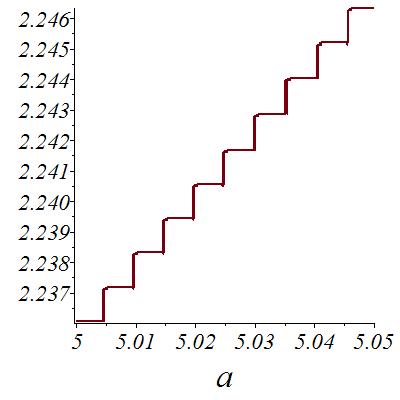}\hss} 
\bigskip

It is worth noticing that, when it fails to compute a closed form
solution, Maple's \texttt{solve} function expresses a result with the
\texttt{RootOf} function, that has a well defined solution.
\smallskip

{\small\parskip=0pt\parindent=0pt

\Mapinp\ $c := solve(x^3+exp(x) = b, x)$; $solve(x^2 = b, x)$\HFB
\Cale\Cale\Cale\Cale  {\color{FOB} $c := RootOf({\rm e}^{\_Z}+\_Z^3-b)$}\HFB 
\Cale\Cale\Cale\Cale  {\color{FOB} $\sqrt{b}, \sqrt{b}$}\HFB 
\Mapinp\ $\Diff(c, b), \Diff(RootOf(\_Z^2-b), b)$\HFB
\Cale\Cale\Cale\Cale  {\color{FOB} $\frac{1}{(exp(RootOf({\rm e}^{\_Z}+\_Z^3-b))+3*RootOf({\rm e}^{\_Z}+\_Z^3-b)^2)},\> \frac{1}{(2*RootOf(\_Z^2-b))}$}\HFB 
\Mapinp\ $apply(eval(\Qt \Diff/RootOf\Qt), P(\_Z, b), b)$\HFB
\Cale\Cale\Cale\Cale  {\color{FOB} $-\frac{(D[2](P))(RootOf(P(_Z, b)), b)}{(D[1](P))(RootOf(P(_Z, b)), b)}$}\HFB 

}
\smallskip

To conclude this topic, one may remark that $i$ iterations of Newton's
method produces a Taylor series up to order $2^{i}$. But in practice
acceptable values for derivatives of higher order may be obtained. The
following procedure applies Newton method to a power series, so that
successive derivatives of the Newton operator may be easily obtained.
\smallskip

{\small\parskip=0pt\parindent=0pt

\Mapinp\ $h := \Proc (a, \eta, c, n, m)$ \Local\ $x, b, d, \epsilon$;
\Global\ $nIter$; $nIter := 0$; $x := c$;\HFB 
\Cale\Cale\Cale \If\ $type(a+\eta+c+n+m, numeric)$ \Then\HFB 
\Cale\Cale\Cale\Cale $b := (a+\epsilon)/x$;\HFB 
\Cale\Cale\Cale\Cale \While\  
$evalb(\eta < abs(subs(\epsilon = 0, convert(series((b-x)/x, \epsilon =
0, n), polynom))))$ \Do\ \HFB
\Cale\Cale\Cale\Cale\Cale $nIter := nIter+1$;\HFB 
\Cale\Cale\Cale\Cale\Cale $x := series((1/2)*x+(1/2)*b, \epsilon = 0, n)$;\HFB 
\Cale\Cale\Cale\Cale\Cale $b := series((a+\epsilon)/x, \epsilon = 0, n)$\HFB 
\Cale\Cale\Cale\Cale \Enddo;\HFB 
\Cale\Cale\Cale\Cale 
$(\Diff(subs(\epsilon^m = d, convert(x, polynom)),
d))*factorial(m)$\HFB 
\Cale\Cale\Cale \Else\ $return$ \Qte $procname(args)$\Qte\HFB 
\Cale\Cale\Cale \Fi\HFB 
\Endproc\HFB
\Mapinp\ $h(2., 0.1e-3, 1., 20, 19)$, $nIter$;\HFB
\Cale\Cale\Cale\Cale\Cale  {\color{FOB}$1.141438794 10^{9}$, $3$}\HFB
\Mapinp\ $subs(b = 2., \Diff(b^{(1/2)},b\$19))$\HFB
\Cale\Cale\Cale\Cale\Cale  {\color{FOB}$1.140326912*10^{9}$}\HFB
}

After just $3$ iteration, the derivatives must be correct up to order
$7=2^{3}-1$. Anyway, by chance, the value of the $19^{\rm th}$
derivative is already obtained with a precision of $10^{-3}$.

\includegraphics[width=4cm]{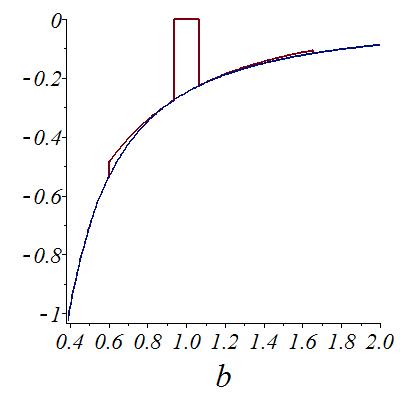}

In such a situation it seems prudent to iterate loops at least one
more time or even $\log_{2}(e)$ if derivatives of order $e$ are to be
computed. In the last figure, acceptable values of the second
derivatives were obtained with a single iteration. Adding extra
iterations is expressed by the old recipe: \textit{one always iterate
  a loop $13$ times!}

\subsection{Integrators and differential equations}\label{Diff-eq}

It is amazing that the requested formulas seem to have left no precise
record in the history of mathematics. We may find some of them,
treated as ``weel known'', in some posthumous manuscript of
Jacobi\cite{Jacobi1}, but as late as 1919, Ritt published a paper
\cite{Ritt1919}\footnote{This paper mentions a previous work of Major
  Forest Moulton, whom he met when he worked with Oswald Veblen for
  the U.S. artillery at Aberdeen Proving Ground during First World
  War.}  on the special problem of differentiating the solution of a
single order one equation $y'=f(y,t,c)$ with respect to the parameter
$c$. One may also mention some works by Bliss
\cite{Bliss1905,Bliss1918} or
Grönwall\cite{Gronwald1919}\footnote{Gilbert Ames Bliss and Thomas
  Hakon Gronwall worked also with Veblen at Aberdeen. The question of
  differentiation with respect to initial conditions is of special
  interest in artillery as a gunner can only play with them. The
  influence of parameters such as projectile mass that cannot be
  exactly matched is also to be studied and compensated e.g. for 155mm
  ammunition using extra propelant bags.}  on the same subject.
\bigskip

Assume that the solution $f_{j}(x_{1}, \ldots, x_{m},\theta_{1},
\ldots, \theta_{s})$ of an
ODE or PDE system 
\begin{equation}\label{PDEs}
P_{i}\left(\frac{\partial^{|\alpha|}f_{j}}{\partial
  x_{1}^{\alpha_{1}}\cdots\partial
  x_{m}^{\alpha_{m}}}(x,\theta),\theta_{1}, \ldots, \theta_{s}\right)=0
\end{equation}
 depending on
parameters $\theta$, completed with initial or boundary conditions
\begin{equation}\label{PDE_limit}
Q_{k}\left(\frac{\partial^{|\alpha|}f_{j}}{\partial x_{1}^{\alpha_{1}}\cdots\partial x_{m}^{\alpha_{m}}}(X(x,\theta),\theta),\theta\right)=0
\end{equation}
 is unique and that it is differentiable with respect to the
 parameters $\theta_{\ell}$. For instance, if $m=2$, one may have
 $X_{1}(x,\theta)=x_{1}$ and $X_{2}(x,\theta)=0$ or
 $X_{2}(x,\theta)=\theta_{1}$ to describe initial condition on an line
 passing through the origin or depending on the parameter
 $\theta_{1}$. One may also use
 $X_{i}=x_{i}/\sqrt{x_{1}^{2}+x_{2}^{2}}$, defined on
 $\R^{2}\setminus\{(0,0)\}$ for initial conditions on the unit circle, etc.

Then, the basic properties of derivation impose that
\begin{equation}\label{PDEs_diff}
\frac{\partial}{\partial
  \theta_\ell}P_{i}\left(\frac{\partial^{|\alpha|}f_{j}}{\partial
  x_{1}^{\alpha_{1}}\cdots\partial x_{m}^{\alpha_{m}}}(x,\theta)\right)
=\sum_{j,\alpha}\frac{\partial P_{i}}{\partial \frac{\partial^{|\alpha|+1}f_{j}}
  {\partial x_{1}^{\alpha_{1}}\cdots\partial
    x_{m}^{\alpha_{m}}}}\frac{\partial^{|\alpha|+1}f_{j}}{\partial
  x_{1}^{\alpha_{1}}\cdots\partial x_{m}^{\alpha_{m}}\partial
  \theta_{\ell}}+\frac{\partial P_{j}}{\partial \theta_{\ell}}= 0
\end{equation}
and this system is to be completed with the derivatives of boundary conditions
\begin{equation}\label{PDE_limit_diff}
\begin{array}{l}
\frac{\partial}{\partial
  \theta_{\ell}}Q_{k}\left(\frac{\partial^{|\alpha|}f_{j}}{\partial
  x_{1}^{\alpha_{1}}\cdots\partial
  x_{m}^{\alpha_{m}}}(X(x,\theta),\theta)\right)=\\
\hbox{\quad\quad}\sum_{j,\alpha} \frac{\partial P_{i}}{\partial
  \frac{\partial^{|\alpha|}f_{j}} 
  {\partial x_{1}^{\alpha_{1}}\cdots\partial
    x_{m}^{\alpha_{m}}}}
\frac{\partial^{|\alpha|+1}f_{j}}{\partial
  x_{1}^{\alpha_{1}}\cdots\partial
  x_{m}^{\alpha_{m}},\theta_{\ell}}(X(x,\theta),\theta)\\
\hbox{\quad\quad}+\sum_{j,\alpha,\mu} \frac{\partial P_{i}}{\partial
  \frac{\partial^{|\alpha|}f_{j}} 
  {\partial x_{1}^{\alpha_{1}}\cdots\partial
    x_{m}^{\alpha_{m}}}}
\frac{\partial^{|\alpha|+1}f_{j}}{\partial
  x_{1}^{\alpha_{1}}\cdots\partial^{\alpha_{\mu}+1} x_{\mu}\cdots\partial
  x_{m}^{\alpha_{m}},\theta_{\ell}}(X(x,\theta),\theta)\frac{\partial
  X_{\mu}}{\partial \theta_{\ell}}+
\frac{\partial Q}{\partial \theta_{\ell}}=0.
\end{array}
\end{equation}
A complete study in the partial differential case would exceed the
ambitions of this paper. One may refer to
\cite{Tijskens2002} for more details and examples in this case.
In the case of ODE, one may also refer to Dickinson \textit{et al.}
\cite{Dickinson1976} and \cite{Elsheikh2015} for implicit differential
algebraic equations. See also Estévez Schwarz \textit{et al.}
\cite{Estevez2015} in the case of singularities.
For generalized ODEs, one may refer to Slavík \cite{Slavik2013}

\bigskip

In the case of ODEs, the following theorem synthetizes the rules to apply.

\begin{theorem}\label{rules}
Let us consider a parametrized system of ordinary differential equations
\begin{eqnarray}\label{sys-edo}
x_{i}'&=&f_{i}(x,t,\theta)\quad 1\le i\le n;\\ 
\label{init_cond}x_{i}(h(\theta))&=&g_{i}(\theta)\quad 1\le i\le n,
\end{eqnarray}
where the functions $f_{i}$, $g$ and $h$ are defined and $\cC^{r}$,
$r\in\N\cup\{\infty\}$, on some open set $V$.

i) There exists a maximal open set $U\in\R^{2}$ and a function
$x:U\mapsto\R^{n}$ such that:\hfill\break
a) $\pi_{2}(V)=U$, where $\pi_{2}(t,\theta)=\theta$;\hfill\break
b) $\forall\theta\in V$ $(h(\theta),\theta)\in U$;\hfill\break
c) $\forall\theta\in V$ $x(\cdot,\theta)$ is a solution of
the differential system~(\ref{sys-edo}) with initial
conditions~(\ref{init_cond}).

ii) The function $x$ is $\cC^{r}$ on $U$.

iii) The derivative functions
$x_{i,\theta}(\partial x_{i}/\partial\theta)$ are solution of the
system
\begin{eqnarray}\label{eqn-derivative}
x_{i,\theta}'&=& \sum_{i=1}^{n}\frac{\partial f_{i}(x,t,\theta)}{\partial
  x_{i}}x_{i,\theta} + \frac{\partial f_{i}(x,t,\theta)}{\partial \theta};\\
\hbox{\quad}\hbox{with initial conditions:}\nonumber\\
\label{init-derivative}x_{i,\theta}(h(\theta))&=& 
\frac{\partial g_{i}}{\partial \theta} - 
f_{i}(x,t,\theta)\frac{\partial h_{i}}{\partial \theta}.
\end{eqnarray}
\end{theorem}
\begin{Proof}
Assuming the functions $x_{i}(t,\theta)$ to exist and be $\cC^{r}$ on
$[a,b]\times[c,d]$, the equations~\ref{eqn-derivative} are similar to
equations~\ref{PDEs_diff} and are obtained by straightforward
differentiation of the initial system~\ref{sys-edo}
$$
\begin{array}{ccl}
&x_{i}(t,\theta)&=f_{i}(x(t,\theta),t,\theta)\\
\Longrightarrow &\frac{\partial
    x_{i}(t,\theta)}{\partial\theta}&=
    \frac{\partial f_{i}}{\partial x}\frac{\partial
      x(t,\theta)}{\partial\theta}
      +\frac{\partial f_{i}}{\partial x}\frac{\partial\theta}.
\end{array}
$$
Differentiating the initial
conditions~\eqref{init_cond} gives
$$
\frac{\partial x_{i}(h(\theta),\theta)}{\partial t}\frac{\partial
  h(\theta)}{\partial\theta} + \frac{\partial
    x_{i}(h(\theta),\theta)}{\partial\theta}=\frac{\partial
  g_{i}(\theta)}{\partial\theta}.  
$$
  Then, replacing $\partial x_{i}/\partial t$ by its value
  $f_{i}(x(t,\theta),t,\theta)$, one gets
  formula~\eqref{init-derivative}.

So, the main point is to prove the existence and differentiability of
functions solution of the system~(\ref{eqn-derivative}) with initial
conditions~(\ref{init_cond}).  The existence is a classical result,
that may be easily achieved using, \textit{e.g.}, Picard's
operator. To prove that the functions $x_{i}(t,\theta)$ are indeed
$\cC^{r}$, one may use Picard's operator again. For simplicity, we may
use
$y_{i}(t,\theta):=x_{i}(t+h(\theta)-h(\theta_{0})+t_{0},\theta+\theta_{0})$,
so that studying differentiability of $x_{i}$ at $(t_{0},\theta_{0})$
is reduced to studying that of $y_{i}$ at $(0,0)$. Let
$C:=\max_{i}\max(\sup_{(z,t,\theta)\in[-\epsilon,\epsilon]^{n+2}}(\partial
f/\partial y_{i},\partial f/\partial\theta)(z+g,t,\theta),
\sup_{[-\epsilon,\epsilon]}\partial g_{i}/\partial\theta)$ and
$y_{[0],i}:=g_{i}(\theta)$. We define Picard's operator by
$P(\phi)(t,\theta):=\phi(0,\theta)
+\int_{0}^{t}f(\phi(\tau,\theta),\theta)\td\tau$ and
$y_{[s]}:=P^{r}(y_{[0]})$. An easy recurence shows that for
$|t|<\ln(\epsilon)/C$ and $|\theta-\theta_{0}|\le\epsilon$, we have $\forall
s\in\N$ $y_{[s]}(t,\theta)<\epsilon$ and
$\|y_{[s+1]}(t,\theta)-y_{[s]}(t,\theta)\|_{\infty}|\le
(Ct)^{s+1}/(s+1)!$. This shows that the sequence
$\|y_{[s]}(t,\theta)\|_{\infty}$ converges uniformly. If the same way, under
the same hypotheses, $\forall q\le r$ $\|\partial^{q} y_{[s]}(t,\theta)/(\partial
\theta)^{q}\|_{\infty}<\epsilon C^{q}$ and
$\|\partial^{q} y_{[s+1]}(t,\theta)/(\partial\theta)^{q}-\partial^{q} 
y_{[s]}(t,\theta)/(\partial\theta)^{q}\|_{\infty}|\le
C(Ct)^{s+1}/(s+1)!$, showing that the sequence $\|\partial^{q}
y_{[s]}(t,\theta)/\partial\theta^{q}\|_{\infty}$ converges uniformly
too. As both the
sequence of functions and of derivatives converge uniformly, the limit
of the functions $y_{[s]}$ is differentiable and its derivatives are
the limit of their derivatives.

To conclude the proof, one just needs now remark that
$x_{i}(t,\theta)=y_{i}(t-h(\theta),\theta)$, so that $(\partial
x_{i}/\partial\theta)(h(\theta_{0}),\theta_{0})=(\partial
y_{i}/\partial\theta)(h(\theta_{0}),\theta_{0})
-(\partial h/\partial\theta)(\theta_{0})(\partial y_{i}\partial
t)(h(\theta_{0}),\theta_{0})$, thus recovering
fomula~\eqref{init-derivative}. 
\end{Proof}

\subsection{Higher order derivatives}

The idea may be iterated, as many times as needed and possible,
according to the differentiability order of the functions defining the
system and initial equations. One may also consider partial
derivatives with respect to an arbitrary number $s$ of parameters. The
only point is to notice that, obviously, $\partial^{2}
x/\partial\theta_{1}\theta_{2}$ $=$ $\partial^{2}
x/\partial\theta_{2}\theta_{1}$. So, whenever possible, one should try
to avoid useless computation instead of applying twice the gradient
operator, provided that the extra implementation work remains
acceptable.

The two formulations are mathematically equivalent, but the numerical
computations will not be the same. An optimized choice would require extra
investigations. Comparing the two results may be a test for strategies
and integrators.

In the generic situation, in order to compute
$\partial^{|\alpha|} y/\prod_{j=1}^{s}\partial \theta_{j}^{\alpha_{j}}$, one
needs to compute all partial derivatives  $\partial^{|\beta|}
y/\prod_{j=1}^{s}\partial \theta_{j}^{\beta_{j}}$ with $\forall j$
$\beta_{j}\le\alpha_{j}$. In other words, if one does not compute the
derivative for $\beta$, one will be unable to compute the one for
$\beta+\gamma$, $\gamma\in\N^{s}$. Computing partial derivatives up to
an arbitrary order, one need not compute them all, but the subset must
be such that the multi-index $\alpha$ are in the complementary set of
an additive \textit{e-set} or \textit{monoideal}, \textit{i.e.} sets
stable by addition of any element of a monoide.
\medskip

Computing high order power series solutions of solutions of ODEs may
be efficiently done using Newton's operator \cite{Bostan2007b}, which
is asymptotically optimal. However
van der Hoeven's method proves more efficient\cite{Hoeven2002} in the
range of order of practical interest.

\subsection{Implementation in Maple and examples} 

We tried to put into practice the theoretical recipes described above,
with the Maple package \href{http://www.lix.polytechnique.fr/~ollivier/AD_Web/D_ODE_tools/D_ODE_tools.mpl}{\texttt{D\_ODE\_tools}} that is
  illustrated with some \href{http://www.lix.polytechnique.fr/~ollivier/AD_Web/D_ODE_tools/Test_D_ODE_tools.html}{examples}.
The basic principle is easy, but to put it in practice may be tedious
and require some effort of style to produce a usable result. At first
sight, the formulae describing the new differential equations and
initial conditions just have to be applied. This would be easy using 
a representation of $x$ as a function of $t$ and the parameters
$\theta$. But, Maple's function \texttt{dsolve} only accepts functions
of one single argument. So, we cannot represent $\partial^{2}
x/\partial\theta_{1}\theta_{2}$ by
\texttt{diff(x(t,theta[1],theta[2]),theta[1],theta[2])}. We need to
imagine some other representation, like
\texttt{x[theta[1]*theta[2]](t)}. Without getting into tedious
details, the greatest part of the work is a carefull analysis of the
internal representation of derivatives in Maple, once faced to the
absence of some ready-made tools to extract elementary information
such as partial or total orders, name of the base function and
independent variables.

The implementation provided here allows to compute derivatives of
arbitrary order with respect to the parameters, that may also appear
in the initial conditions. One may describe the set of derivatives by
excluding some multiples of a given derivation. For this, one uses the
\texttt{Groebner} package of Maple.

One encounters extra limitations of various types. For example,
\texttt{dsolve} may manage to find a symbolic general solution, but is
is impossible to use this function with symbolic boundary conditions
for different values of the independent variable. Our implementation
allows such specifications.
Although our focus is on numerical resolution, systems with closed
form solutions, besides their intrinsic interest, provide easy tests
for the exactness of such an implementation.

\subsubsection{Identifiability}

On may refer to Walter \cite{Walter1982} for a general introduction to
the notion.  We condider a \textit{structure}, \textit{i.e.} a
parametric system of state space equations
\begin{eqnarray}
x_{i}'&=&f_{i}(x,\theta,t),\>1\le i\le n,\> \theta=(\theta_{1}, \ldots,
\theta_{s})\\
x_{i}(0)&=&c_{i}\\
y_{j}&=&g_{j}(x,\theta)\quad 1\le j\le q,
\end{eqnarray}
where the $y_{j}$ are assumed to be \textit{outputs}, or
\textit{mesured quantities}. A \textit{model} of the structure is
obtained by specializing parameters. Then, a model is
\textit{(locally observable} if one can compute a (locally) unique
value of the state functions $x_{i}$ knowing the outputs $y_{j}$. It
is  \textit{(locally identifiable} if one can compute a (locally) unique
value of the parameter vector $\theta$ knowing the outputs
$y_{j}$. \textit{Structural} observability or identifiability means
that almost all models are observable or identifiable. If the
functions $f_{i}$ and $g_{j}$ are polynomial or rationnal, the vector
of parameters corresponding to locally observable or identifiable
models is empty or dense. So, testing these structural properties  may
be done by finding a single observable or identifiable model. This may
be done by symbolic computation. Biologists are likely to feel more
comfortable with a numerical test, that may be done with the same
program they use for their simulations and requires no exotic
algebraic knowledge, besides matrix ranks and determinants.
But they are likely to ignore that one may compute $\partial y_{j}/\partial
\theta_{\ell}$ in a better way than using finite differences.

The most basic test is to choose $s+q$ times $t_{k}$ and to check if
the determinant
$$
\left|
\begin{array}{cc}
\frac{\partial y_{j}}{\partial c_{i}}(t_{k})&
\frac{\partial y_{j}}{\partial \theta_{\ell}}(t_{k})
\end{array}
\right|
$$
is non $0$. Then the system is both locally observable and identifiable.

One must remark that numerical computation can show that the
determinant does not vanish, but not that it is $0$. In such a case,
one will get a small result, depending on the precision. Moreover,
even the vanishing of the determinant does not prove that the system
is not observable and identifiable: there is a probability to choose
unfortunate values for the $t_{k}$ that are precisely roots of this
determinant. Using exact computation, one can obtain a probabilistic
test, meaning that the system can be shown to be non observable with a
small possibility of error. See Sedoglavic \cite{Sedoglavic}. This
method uses fast power series computation using Newton's operator
described in \cite{Bostan2007b}. 

An example of some HIV model due to Perelman \textit{et al.} is given
as an
\href{http://www.lix.polytechnique.fr/~ollivier/AD_Web/IDENTIFIABILITY/Perelson_3_numeric.html}{illustration}.
\medskip

See Schumann-Bischoff \cite{Schumann2015} \textit{et al.} for the use
of AD in parameter identification.

\subsection{Delay systems}

Differential systems with delays enter easily in our general
setting. Latest versions of Maple allow to solve them with
\texttt{dsolve}. Among important issues that are linked with the
differentiation of delay systems with respect to their parameters,
including the delay itself, is the practical online computation of an
unknown delay. See e.g.

The rules are easy to generalize to this setting.
Assume a differential system with delays is given by

$$
x_{i}'(t)=f_{i}(x(t),x(t_{h}),t),
$$
then its derivative is a solution of the system
$$
x_{h,i}'(t)=\sum_{j=1}^{n}\frac{\partial f_{i}}{\partial
  x_{j}(t)}x_{h}(t)+\sum_{j=1}^{n}\frac{\partial f_{i}}{\partial
  x_{j}(t-h)}(x_{h,j}(t-h)-f_{i}(x(t),x(t_{h}),t). 
$$ This is easily generalized to many delays, or delays depending on
parameters, the $x_{i}$ or the time. See \cite{Hartung2006} for more
details in the case of state-dependent neutral functional differential
equations.

\subsection{Sequences}

In many cases, sequences are iterated in order to converge to a fixed
point, a problem that we have already considered. But it may happen
that they possess a different meaning and must be iterated a precise
number of times. Such an issue cannot be solved by inspecting a low
level program.

The following example shows that computing the derivative may
sometimes be almost impossible, even if the iteration of the function
is easily computed:
$$
f(x,n)=(10^{n}x, n+1).
$$
The value of $f^{21}(0,-10)$ is $(0,11)$ and $\partial f^{21}/\partial
x(0,-10)$ is $1$. But the accurate computation of this value with
floating point arithmetic would require an unusual precision. See also
Beck \cite{Beck1994b}.

\subsection{Conditionals and discontinuities in EDOs}

We have already discussed the case of continuous functions define by
conditionals. The case of discontinuous functions reduces at first
sight to the impossibility of computing derivatives at
discontinuities. The case becomes harder when we integrate such
functions. The derivative of the Heaviside function is $0$ on
$\R^{\ast}$ and undefined at $0$. But $\int_{-\infty}^{y}
\hbox{\it Heaviside}(x-a) dx=(y-a)\hbox{\it Heaviside}(y-a)$, so that
its derivative with 
respect to $a$ is defined and not $0$ for $y>a$. 

Such issues cannot be handeled in Maple by the \texttt{piecewise}
function but the derivative of the \texttt{Heaviside} function
involves the \texttt{Dirac} function and allows correct computation.

It is however almost impossible to adapt the idea in some numerical
setting and most of the time the problem remains unsolved. The best
compromise is to approximate the Heaviside function with
\texttt{atan(ax)}, as done in \texttt{Diffedge},
or \texttt{erf(ax)}. The choice of the parameter $a$ and of the
function to be substituted to Heaviside may be a delicate issue for
which we found no proper reference. If $a$ is too small, the
approximation of the Heaviside function is poor. If too great,
integrators will face difficulties to integrate a stiff equation. We will see in the next section
how the use of ``\texttt{events}'' when solving differential equation
may offer an alternative and permit some comparisons.
\medskip

A few \href{http://www.lix.polytechnique.fr/~ollivier/AD_Web/DIRAC/Dirac.html}{Maple experiments} illustrating the topic are provided.

\subsection{Lagrangian}

Mechanical systems whose equations may be easily computed using the
Lagrangian formalism deserve a special attention. Let us recall that
the Lagrangian is expressed by
$\cL(x'(t),x(t),t):=E_{k}(x'(t),x(t),t)-E_{p}(x(t),t)$, where $E_{k}$ 
denotes kinetic energy and $E_{p}$ potential energy, and that we have
\begin{equation}\label{eq-lagrangien}
\frac{\td}{\td t} \frac{\partial\cL}{\partial
  x_{i}'}=\frac{\partial\cL}{\partial x_{i}}, 
\end{equation}
which expresses the minimality of
$\int_{t_{1}}^{t_{2}}\cL(x'(t),x(t),t)dt$, for any two fixed times
$t_{1}$ and $t_{2}$, the values $x(t_{1})$ and $x(t_{2})$
being fixed too.

Then, if we have a Lagrangian depending on parameters $a_{k}$, we may
first compute a normal form from equations (\ref{eq-lagrangien}),
which requires that the Hessian $|\partial^{2}\cL/\partial
x_{i}'\partial x_{j}'|$ does not vanish, and then compute the
derivatives of $\partial x(t)/\partial a_{k}$ using the method
developped in subsection~\ref{Diff-eq}. An alternative is to stay in
the Lagrangian formalism, using the derivative of the Lagrangian with
respect to $a_{k}$:
$$
\frac{\td}{\td a_{k}}\cL=\sum_{i=1}^{n}\frac{\partial\cL}{x_{i}'}\frac{\partial
  x_{i}'}{\partial a_{k}} + \sum_{i=1}^{n}\frac{\partial\cL}{x_{i}}\frac{\partial
  x_{i}}{\partial a_{k}} + \frac{\partial\cL}{\partial a_{k}}.
$$
Such formulas may me iterated for higher derivatives.
\bigskip

More interesting is the hability to use the Lagrangian formalism with
discontinuous terms of energy in order to model shocks, under the
assumption that the total energy is preserved. We need here the
assumption that the kinetic energy $E_{k}$ is quadratic with respect
to derivatives $x_{i}'$: $E_{k}=(x')^{\rm t}A(x)(x')$, where $A(x)$ is
an \textem{invertible}\footnote{Without invertibility, we cannot
  obtain explicit equations from the Lagrangian.} $n\times n$
matrix. We assume that $E_{p}=E_{p,1}$ if $f(x,t)\ge0$ and
$E_{p}=E_{p,2}$ if $f(x,t)<0$. By the principle of least action, the
following integral must be minimal:
$$
\int_{t_{0}}^{t_{1}}\cL(t)\td t+\int_{t_{1}}^{t_{2}}\cL(t)\td t,
$$
where $f(x(t_{1}),t_{1})=0$. We recall that $x(t_{0})$ and $x(t_{2})$
are fixed. We further assume that 
$$
\sum_{i=1}^{n}\frac{\partial f}{\partial x_{i}}x_{i}'+\frac{\partial f}{\partial t}>0.
$$
In the sequel, we omit sum signs to alleviate formulas.  The
minimality of the integral implies that, denoting by $\delta x_{i}$ a
small variation,
\begin{equation}
\int_{t_{0}}^{t_{1}}\frac{\partial\cL}{\partial
  x_{i}'}\delta x_{i}'+\frac{\partial\cL}{\partial x_{i}}\delta x_{i}\td
t+\int_{t_{1}}^{t_{2}}
\frac{\partial\cL}{\partial
  x_{i}'}\delta x_{i}'+\frac{\partial\cL}{\partial x_{i}}\delta x_{i}\td
t+(\cL(t_{1}^{-})-\cL(t_{1}^{+})\delta t_{1}=0,
\end{equation}
where $\cL(t_{1}^{\pm})$ denote the left and right limits at time
$t_{1}$. Integrating by parts, we get:
\begin{multline}
\int_{t_{0}}^{t_{1}}\left(-\frac{d}{\td t}\frac{\partial\cL}{\partial
  x_{i}'}+\frac{\partial\cL}{\partial x_{i}}\right)\delta x_{i}\td t
+\int_{t_{1}}^{t_{2}}\left(-\frac{d}{\td t}\frac{\partial\cL}{\partial
  x_{i}'}+\frac{\partial\cL}{\partial x_{i}}\right)\delta x_{i}\td t\\
+\frac{\partial\cL}{\partial x_{i}'}(t_{1}^{-})\delta x_{i}(t_{1}^{-})
-\frac{\partial\cL}{\partial x_{i}'}(t_{1}^{+})\delta x_{i}(t_{1}^{+})
+\left(\cL(t_{1}^{-})-\cL(t_{1}^{+})\right)\delta t_{1}=0.
\end{multline}
As we have 
$$
\frac{d}{\td t}\frac{\partial\cL}{\partial
  x_{i}'}=\frac{\partial\cL}{\partial x_{i}},
$$
the condition at $t_{1}$ becomes
\begin{equation}
\frac{\partial\cL}{\partial x_{i}'}(t_{1}^{-})\delta x_{i}(t_{1}^{-})
-\frac{\partial\cL}{\partial x_{i}'}(t_{1}^{+})\delta x_{i}(t_{1}^{+})
+\left(\cL(t_{1}^{-})-\cL(t_{1}^{+})\right)\delta t_{1}=0,
\end{equation}
that must stand for any small variation $\delta x$, $\delta
t_{1}$. For simplicity, we may assume to have chosen coordinates
$x_{i}$ in such a way that $\td f(x(t_{1}),t_{1})=C\td x_{n}$, meaning
that $\td x_{i}$, $^1\le i<n$ are coordinates on the hyperplane
tangent to the hypersurface $H$ defined $f(x(t_{1}),t_{1})=0$. With
$A(x(t_{1}),t_{1})=(a_{i,j})$, we may furthermore
assume that $a_{i,n}=a_{n,i}=0$ if $i\neq n$.

With $\td t_{1}=0$, we get 
\begin{equation}
\frac{\partial\cL}{\partial x_{i}'}(t_{1}^{-})
=\frac{\partial\cL}{\partial x_{i}'}(t_{1}^{+}),\quad \hbox{for}\>
1\le i<n,
\end{equation}
meaning that 
\begin{equation}
x_{i}'(t_{1}^{-})
=x_{i}'(t_{1}^{+}),\quad \hbox{for}\>
1\le i<n.
\end{equation}
The component of the speed tangent to $H$ is preserved.

As the trajectory is continuous, we have 
\begin{equation}
\delta
x_{i}(t_{1}^{-})+x_{i}'(t_{1}^{-})\delta t_{1}=\delta
x_{i}(t_{1}^{+})-x_{i}'(t_{1}^{+})\delta t_{1}
\end{equation} 
with
\begin{equation}
\frac{\partial f}{\partial x_{i}}\delta
x_{i}(t_{1}^{\pm})\mp\left(\frac{\partial f}{\partial
  x_{i}}x_{i}'(t_{1}^{\pm})+\frac{\partial f}{\partial t}(t_{1})\right)\delta t_{1}=0.
\end{equation}
We may now consider the case $x_{i}(t_{1}^{-})+x_{i}'(t_{1}^{-})\delta t_{1}=\delta
x_{i}(t_{1}^{+})-x_{i}'(t_{1}^{+})\delta t_{1}=0$ for $1\le i<n$. Let 
$$
v:=-\frac{\partial f}{\partial t}/\frac{\partial f}{\partial x_{n}}.
$$
It corresponds to the speed of the hyperplane $H$. We have
$$
\delta x_{n}(t_{1}^{\pm})= \pm(x_{n}'(t_{1}^{\pm})-v)\delta t_{1},
$$
so that we obtain
\begin{equation}
2a_{n,n}x_{n}'(t_{1}^{-})(x_{n}'(t_{1}^{-})-v)+a_{n,n}x_{n}'(t_{1}^{-})^{2}+E_{p,2}=
2a_{n,n}x_{n}'(t_{1}^{-})(x_{n}'(t_{1}^{-})-v)+a_{n,n}x_{n}'(t_{1}^{+})^{2}+E_{p,1}
\end{equation}
equivalent to the final equation:
\begin{equation}\label{eq-final}
a_{n,n}(x_{n}'(t_{1}^{-})-v)^{2}+E_{p,2}=a_{n,n}(x_{n}'(t_{1}^{+})-v)^{2}+E_{p,1},
\end{equation}
meaning that the sum of the potential energy and the kinetic energy
associated to the speed relative to the hyperplane $H$ and orthogonal
to it for the norm defined by the matrix $A$ remains constant. In
particular, this implies the conservation of the energy.

Assuming that $a_{n,n}(x_{n}'(t_{1}^{-})-v)^{2}+E_{p,1}-E_{p,2}<0$,
the equation (\ref{eq-final}) has no real solution, so that we have a
rebound. The mobile stays in the half space $f(x(t),t)<0$ and we have
then
\begin{equation}
a_{n,n}(x_{n}'(t_{1}^{+})-v)=-a_{n,n}(x_{n}'(t_{1}^{-})-v).
\end{equation}
\bigskip

The Maple package \texttt{D\_ODE\_tools} contains a function that
computes
\href{http://fr.maplesoft.com/support/help/maple/view.aspx?path=dsolve/numeric/Events}{\texttt{events}}
that represent such a discontinuity, associated to the vanishing of
function $f$. Details of the implementation would be boring. We only
mention that the evaluation of the new values $x_{i}'(t)=\hbox{\it
  foo}_{i}(x')$ is done in sequence, which would produce the erroneous
affectation $x_{2}'=\hbox{\it foo}_{2}(\hbox{\it foo}_{1}(x'),x_{2}',
\dots)$ instead of $x_{2}'=\hbox{\it foo}_{2}(x_{1}',x_{2}', \dots)$,
so that we needed to perform in two steps $z_{i}=\hbox{\it
  foo}_{i}(x')$, and then $x_{i}'=z_{i}$. What is to be stressed is
that AD of high level functions is full of many troubles of this kind
that are highly time consuming: it is unavoidable that the documentation
of a powerfull and versatile function tends to be both quite lengthy
and somehow unprecise, so that many experiments are required.
\bigskip

As an example, the case of a \href{http://www.lix.polytechnique.fr/~ollivier/AD_Web/LAGRANGIAN/Lagrangian.html}{double pendulum}. The integration using
\texttt{events} is compared with an approximation with $\arctan$. One
may note that we need to use some trick to prevent unwanted activation
of an event a short time after a discontinuity has been encountered
and first derivatives given new values.
\bigskip

In our Maple worksheet, many intermediate results are not printed. In
fact, the formulas used in the events description are often (too)
big. One needs tools to replace useless lengthy formulas with SLP
(Straight Line Programms) that compute them. Such issues have been
considered with success for solving algebraic systems, for example
with
\href{http://lecerf.perso.math.cnrs.fr/software/kronecker/index.html}{Kronecker}. See
Giusti \textit{et al.} \cite{Giusti2001}.
\bigskip

Producing new events to compute the derivatives of solutions of ODE
with discontinuities produced by events accoding to the dsolve syntax
boils down from a mathematical standpoint to the case of initial
conditions, already considered in
subsection{Diff-eq}~th.~\ref{rules}. From a practical standpoint, we
prefer to postpone this task, due to the difficulties mentionned
above, and many others of the same kind, among them the necessity to
deal with the many cases of possible ``events'', not all associated to
possible discontinuities.  We hope to go back to with issue in a next
version of this paper.
\bigskip

Glocker \cite{Glocker2004} considers impacts with dissipations,
or multicontact collisions.
We refer to Fetecau \textit{et al.} \cite{Fetecau2003} or Leine
\textit{et al.} \cite{Leine2004} and the
references therin for more details on rigid-body dynamics with
impacts, dry friction, etc.\ 
in a more general setting. 

\subsection{Operational Calculus}\label{Operational-calculus}

Operational calculus is a convenient way to deal with linear
differential equations, popular in many applicational fields,
including control theory. An easy way to develop it on sound bases, avoiding 
difficulties of defining the Laplace transform for some function, was
developped by the Polish mathematician Jan Mikusiński
\cite{Mikusinski1953}, using Titchmarsh convolution theorem
\cite{Titchmarsh1926}: the convolution operator on $\R^{+}$ defines a
domain on continuous functions. So, as $\int f=1\ast f$, the
derivation operator $s$ may be defined as the inverse of $1$ for
convolution, in a purely algebraic way.

Even delays may be modeled in this framework, by using formula
$f(x-h)={\rm e}^{-hs}f$. Differentiation with respect to parameters in
differential equations is straightforward in this setting. If we have
a differential equation decribing a system with output $y$ and input
$u$: $y'= ay + u$, we may translate it as $(s+1)y=u$, so that
$y=u/(s+1)$. This \textem{transfer function} expresses the relation
between the input and the output.

\subsection{Differentiating noisy data}

If the final goal of automatic differentiation is to build enbeded
code to control and optimize the behaviour of a physical process, one
is soon faced with the major difficulty of computing the derivatives
of noisy signal. Finite difference may be acceptable for first
derivatives and low level noises, but one may achieve better accuracy
using the evaluation of derivatives by integration.

The general idea may be traced back to \cite{Lanczos1956}. See also
\cite{Groetsch1998,Shen1999,Shen1999,Rangarajan2005,Dridi2010}. But
this approach is not well suited for online computation as each
evaluation of a derivative at a given point require a new integration. An
other method inspired by Mikusiński's theory has been
designed by Fliess, Sira-Ramirez, Join\dots\  \cite{Fliess2008}.
With this approach, the evaluation of derivatives may be obtained by
continuous integration. However, the quality of the result is best
some time after the integration starts and is subject to some
degradation, so that one needs to restart repeatedly the method, which
is not trivial from a computer programming standpoint.
 See
also \cite{Ollivier2007} for a variant that tries to avoid
reinitialization and may to some extend be used with delay systems.

\section{Diffedge}

\href{http://appedge.pagesperso-orange.fr/diffedge_intro.htm}{Diffedge}
is a differentiation tool designed for systems represented by block
diagrams in Matlab\footnote{About AD in the Matlab environment, see
  \textit{e.g.} \cite{Neidinger2010,Kharche2011}.} Simulink. Its goal
is to provide a result that remains in the same diagram
environment. It was developped by
\href{http://www.appedge.com/english/en_accueil.html}{Appedge} in 2005
\cite{MasseCambois2004,Bastogne2007} and is the result of a long
familiarity with the suject \cite{Gilbert1991}. Optimization often require to
get an analytical expression of the gradient function, which is a non
trivial task in this setting, but then real time optimization tools
may be tested and translated in C/C$^{++}$ code ready for use on
embedded processors, using e.g. Embedded
Coder$^{\mbox{\scriptsize{\textregistered}}}$.

Several ways may be considered, but they do not all allow to handle
models with discontinuities (switch, saturation, etc.) or heterogenous
mathematical representation mixing continuous and discrete time
models. We will illustrate the possibility on some simple examples.

\subsection{A didactic example}

\begin{figure}[!h]
\begin{center}
\includegraphics[scale=0.8]{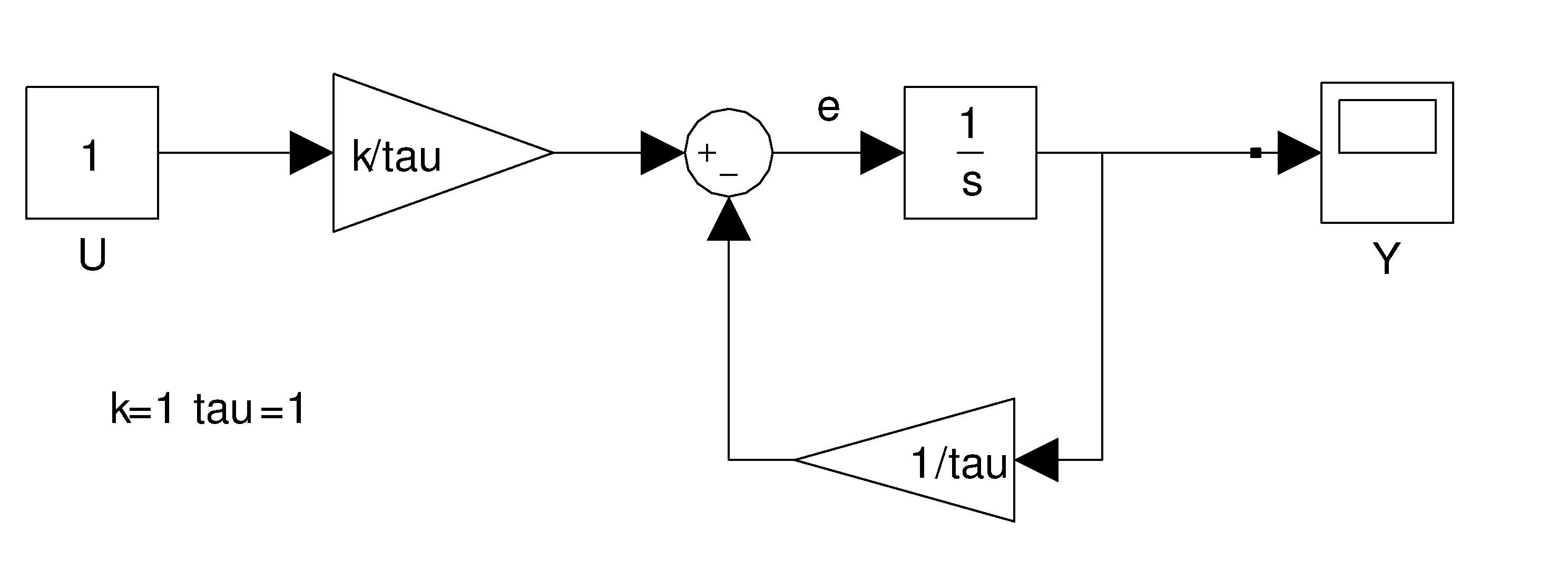}
\caption{Academic example : First order transfert function}
\label{Original1}
\end{center}
\end{figure}

In this first order model (fig.~\ref{Original1}),
we want to compute the derivative with respect to the parameter $\tau$ of
the model.

\subsubsection{Symbolic derivative. Manual method using the transfer
  function}

See subsection~\ref{Operational-calculus} for operational calculus
formalism used here.
Compute the equivalent transfer
function. Write the equation between 
each block and resolve the causality:
\begin{equation}\label{transfer}
\left(
\begin{array}{lcl}
e&=&-\frac{1}{\tau}Y+\frac{k}{\tau}U\\
Y&=&\frac{1}{s}e
\end{array}\right)
\quad
\Longrightarrow\quad Y = \frac{kU}{1+\tau s}
\end{equation}

Then, we have two possibilities. The first is to work in the Laplace
domain.

\begin{equation} \label{transferder}
\frac{dY}{d\tau}=\frac{-ks}{(1+\tau s)^{2}}U.
\end{equation}

The second way is to transform the equation~\ref{transfer} in time
differential equation like (equ.~\ref{timeder}) :
\begin{equation} \label{timeder}
y'=\frac{-y+ku}{\tau} \Longrightarrow \frac{dy'}{d\tau}=
\frac{\left(-\frac{dy}{d\tau}+k\frac{du}{d\tau}\right)\tau +y -ku}{\tau^{2}}
\end{equation}
with
$$
\frac{du}{d\tau}=0.
$$

For theses 2 ways (equ.~\ref{transferder}) and (equ.~\ref{timeder}), it is necessary to rewrite the equations in block
diagram description and to make the time integration to get the
expected result. We can also, when it is possible, compute a finite form
solution (equ.~\ref{tempo}) and derivate this one with
respect to the independent parameters.

All these methods are not useful to keep the block diagram description
but it is better that nothing when the model is simple.
\begin{equation} \label{tempo}
y'=\frac{-y+ku}{\tau}\quad \begin{array}{c}[b]\int\\\longrightarrow\end{array}
\quad y(t)=k\left(1-e^{-t/\tau}\right)u(t) 
\quad \begin{array}{c}[b]d/d\tau\\\longrightarrow\end{array}
\quad \frac{dy(t)}{d\tau}=\frac{kte^{-t/\tau}}{\tau^{2}}.
\end{equation}

\subsubsection{Finite difference with independent parameters} 
Often, this method is
used with optimization algorithm and to compute the derivative with
respect to parameter by finite difference. Nevertheless, it is very
difficult to choose the epsilon value of the parameter. This method
does not work well when we have noise on the input, strongly non
linear equation or when we have discontinuities in the model ( step,
hysteresis, etc) and it is very difficult to embed it in real time.
The finite difference is not generally accurate and the optimization,
even if we use a complex strategy of optimization, does not lead to
the optimal solution due to the choice of epsilon. But all the
engineers use it for small and big models.

\subsubsection{Via computer algebra tools} We can use
\texttt{BlockImporter}{\scriptsize\texttrademark} (which is a Maple
add-on) that allows you to import a Simulink model into Maple and to
convert it into a set of mathematical equations. Thus it is possible
to analyze and derivate the equation in a reliable way. But the work
is not finished, as it is necessary to translate the result back in
the Simulink block diagram environment. It will be impossible to treat the blocks:
discontinuities, logic evens and look-up table in the model,
except if we decompose the model in piecewise continuous
functions. Another hard issue will be closed form integration, which
will not be possible in the general case, and may be very expensive in
time and memory.

\subsubsection{Automatic differentiation of code}
We may try to convert the model in C or Fortran code and then
call appropriate tools. For a short and didactic example, we
use the on-line Automatic Differentiation engine TAPENADE. The routine
model table~\ref{source-code} describes the equation\eqref{transferder}: the dependent output variables
are \texttt{yprime}, \texttt{y} and the independent parameter \texttt{tau}. We have chosen the
option Fortran and differentiated in ``TangenteMode'', we have
obtained the differentiated program in table~\ref{diff-prog}.

\begin{figure}[h!]\label{source-code}
\begin{center}
{\hsize=10cm
\begin{tabular}{|c|}
\hline
\vbox{{\scriptsize 
\begin{verbatim} 
SUBROUTINE MODEL(yprime, y, u, tau, k)
IMPLICIT NONE
DOUBLE PRECISION yprime, y, u, tau, k
yprime = (-y+k*u)/tau
END
\end{verbatim}
}}
\\
\hline
\end{tabular}
}
\caption{Source code}
\end{center}
\end{figure}

\begin{figure}[h!]\label{diff-prog}
\begin{center}
{\hsize=10cm
\begin{tabular}{|c|}
\hline 
\vbox{{\scriptsize 
\begin{verbatim} 
SUBROUTINE MODEL_D(yprime, yprimed, y, yd, u, tau, taud, k)
IMPLICIT NONE
DOUBLE PRECISION yprime, y, u, tau, k
DOUBLE PRECISION yprimed, yd, taud
yprimed = (-(yd*tau)-(y+k*u)*taud) /tau**2
yprime = (-y+k*u)/tau
END
\end{verbatim}
}}
\\
\hline
\end{tabular}
}
\caption{Differentiated program}
\end{center}
\end{figure}

Usually it is not possible to use the result of the automatic
differentiation in real time. We face a lot of problems of
compatibility between libraries, or related to
the structures of languages,
especially when we use triggers\footnote{Triggers are blocks that
  determine times when the trigger is set ``up'' or ``down'',
  corresponding to special actions to be performed.} or enable blocks. Moreover
we don't have access at all the sources of the libraries, which is a
major limitation to build embedded code.

\subsubsection{Graphic derivative AGDM First derivative}
 An another way is to apply
the rules described in the forthcoming paragraph~\ref{AGDM} Rules for the
automatic graphic differentiation Methodology (AGDM). With these
rules, the model (Fig.\ref{Original1}) can be written like in the
following Figure~\ref{ADGM}. The rules are straightforward to
implement.

\begin{figure}[!h]
\begin{center}
\includegraphics[scale=0.7]{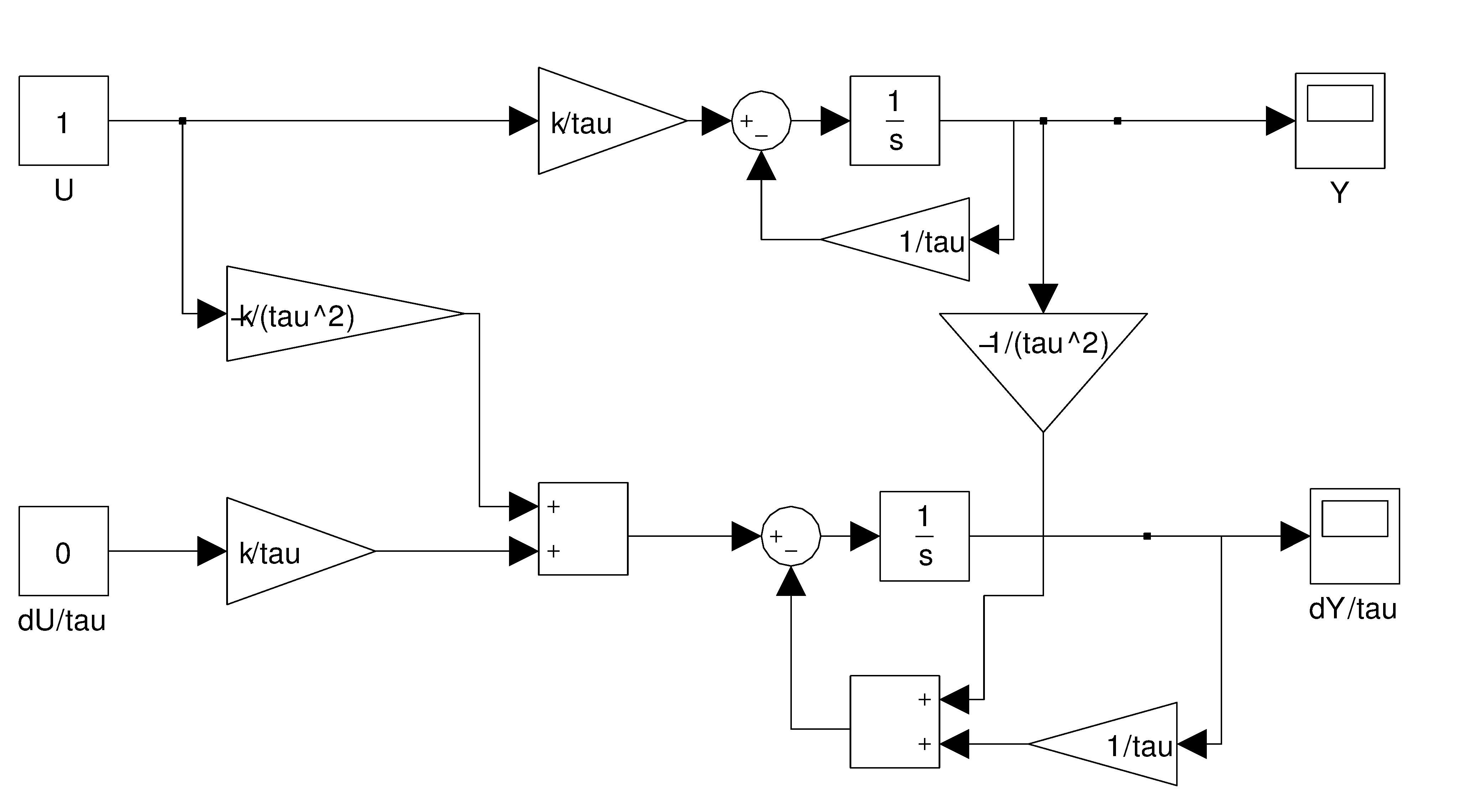}
\caption{ADGM Order $1$}
\label{ADGM}
\end{center}
\end{figure}

\subsection{Computing the Hessian} 
Try to compute the hessian or the second
derivative is very complicated for all the raisons explained before
but also by the size of result too, except if the model is simple.

For example, if we start with the Equation \ref{transfer}, we obtain
the Equation \ref{ddtransfer}. 
\begin{equation}\label{ddtransfer}
\frac{d^{2}Y}{d\tau^{2}}=\frac{2ks}{\left(1+\tau s\right)^{3}}U
\end{equation}
But, the disadvantage which have been
outlined is that we lost, in the condensed formula, the saturation of
integrator. With ADGM it is possible to drive the integrator in the
derivative flow by the saturation value in the original scheme
(Fig.~\ref{Original1}).

\begin{figure}[!h]
\begin{center}
\includegraphics[scale=.9]{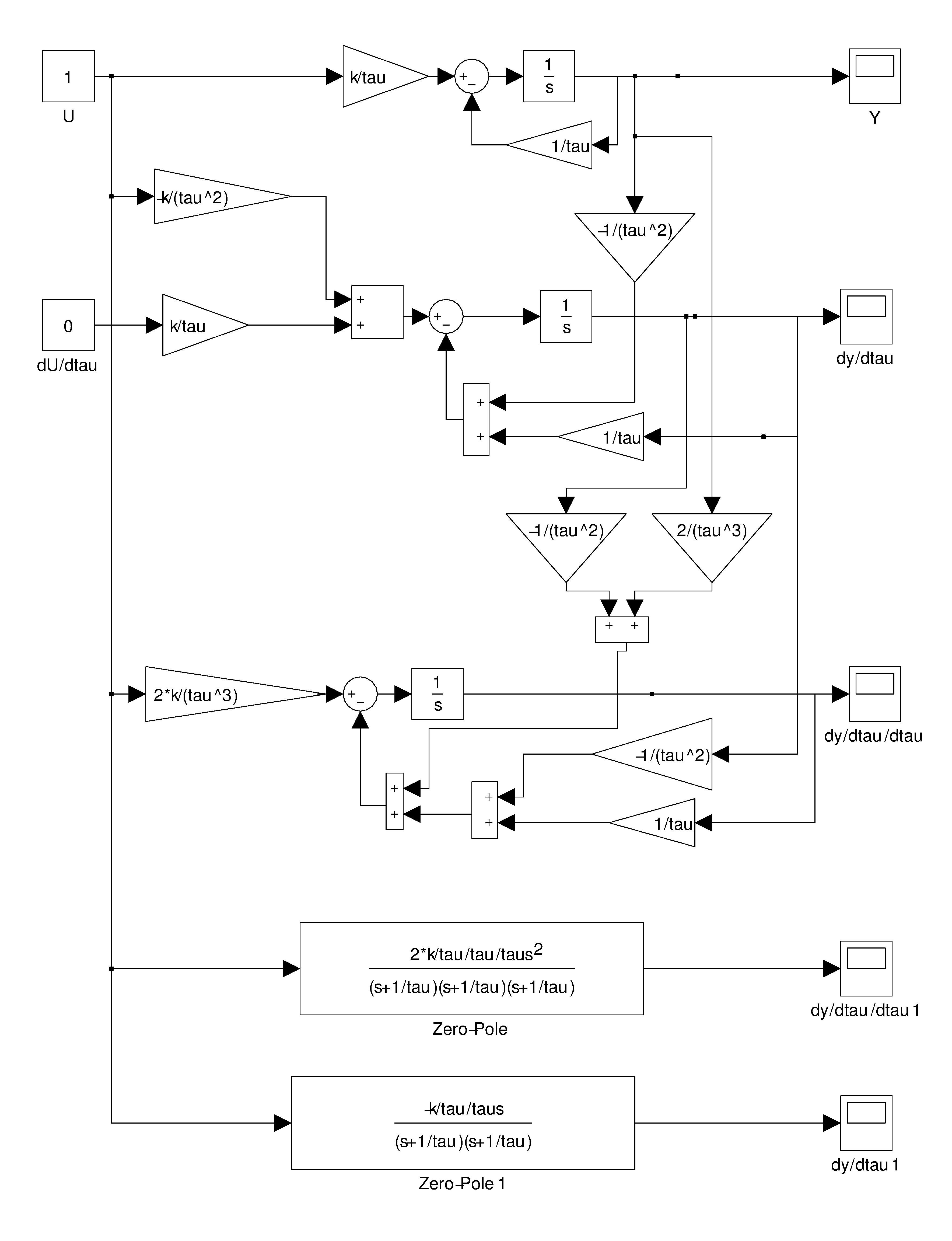}
\caption{Hessien of original scheme ( Order 1)}
\label{Hessien}
\end{center}
\end{figure}

Via ADGM, it is necessary to apply the rules one more time on the
model. The properties of the partial derivative are kept:
$\frac{\partial^{2}Y}{\partial x\partial
  z}=\frac{\partial^{2}Y}{\partial z\partial x}$. However, to use an
helpful automat like Diffedge will be appreciated and we obtain the
Figure \ref{Hessien}. If we compare by numerical simulation ( integration)  the mathematical  form  of ADGM methology and the explicit derivative ( Second order) we obtain exactly the same result. 

\subsection{Rules for Automatic Graphic Differentiation
  Methodology  (AGDM)}\label{AGDM}

The rules are based on the technical field of ``Automatic
Differentiation'' to derive the conditional structures as well as the
structural properties of block (math function, state,...) diagrams and
by application of the formula for derivatives of composite functions.
Seven simple rules of differentiation, including two rules on the
links ( J) and five on the blocks (M) allow to automatize the process
of differentiation which can be manual or by computer using a computer
algebra system like Maple . All these rules can differentiate easily a
model described as pattern blocks (SISO, MIMO, continuous / sampled,
finite state) systems Applying all these rules are possible because we
use the structural property (flow causality, block independent,...) of
bloc diagram representation.  Each block is independent, in other
words it does not depend on other block around it and is
self-standing. This guarantees that all blocks in the block diagram
can be differentiated independently of each other.  Furthermore, we
use the causality property of the block diagram. The derivative flow
propagates like the causality flow.

Finally, with these rules we obtain the block diagram differentiated
with respect to the parameter k. The user benefice is very important
because the derivative model is also a block diagram and like that, we
still have access at all the functionality of Matlab/Simulink
(optimization, real time etc.).  This method is definitively the best
and the most powerful way to get the symbolic derivative instead of
translating the model in equations and then derivate them.  For
understanding these rules, one may look to their application on the
figures (Fig.\ref{Original1}),(Fig.\ref{ADGM}) and
(Fig.\ref{Hessien}).

\subsubsection{Link rules: J1 and J2} 

\begin{figure}[!h]
\begin{center}
\includegraphics[scale=.7]{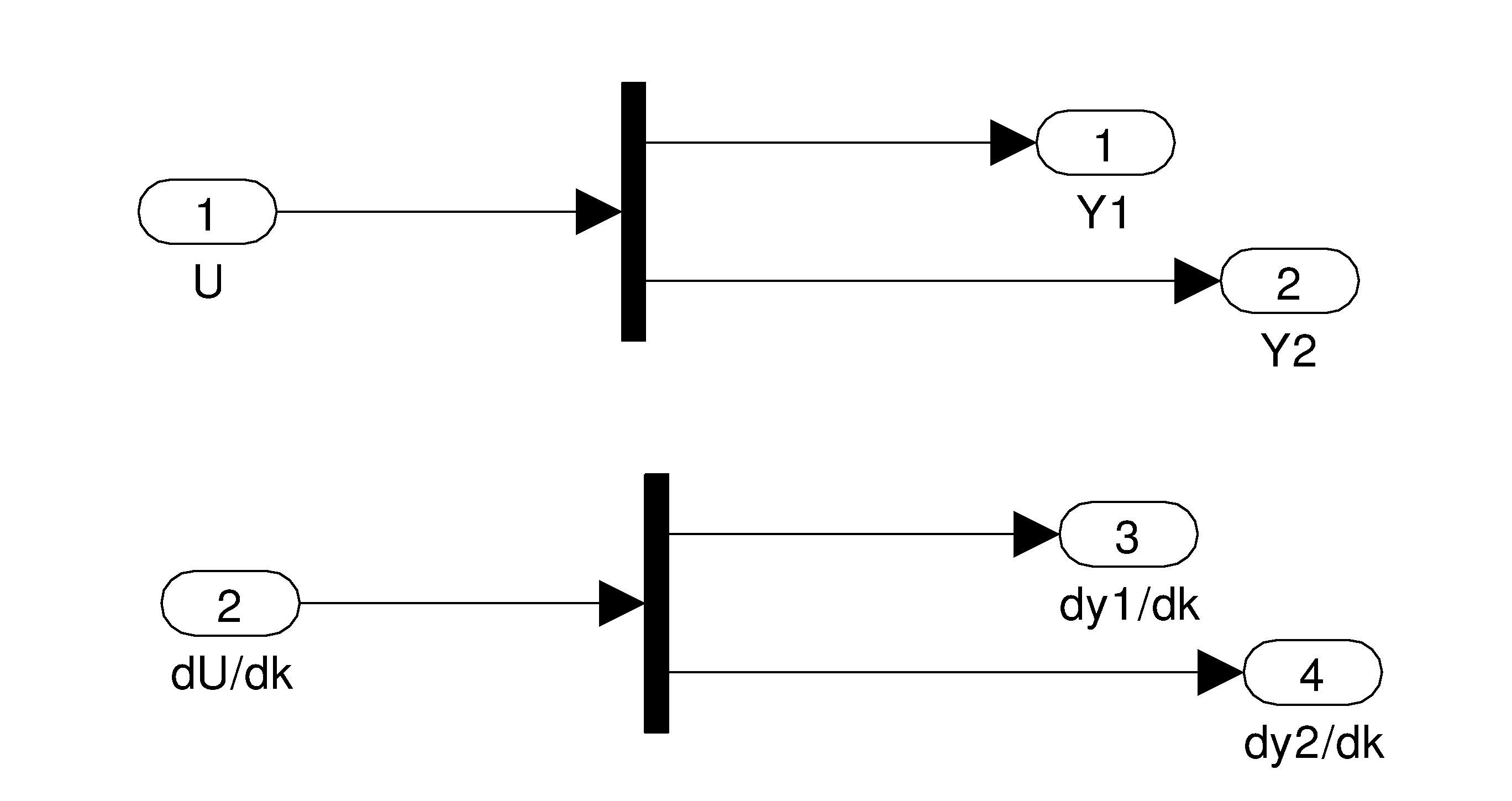}
\caption{J1 rule applied to Mux block}
\label{J1_rule}
\end{center}
\end{figure}

\paragraph{J1}
Whatever the block through by the
differentiated flow, all its outputs will be affected by it. This rule
can be applied for scalar links, vector and matrix. For example
multiplexor, demux, subsystem, etc (see \ref{J1_rule})

\paragraph{J2}

\begin{figure}[!h]
\begin{center}
\includegraphics[scale=.7]{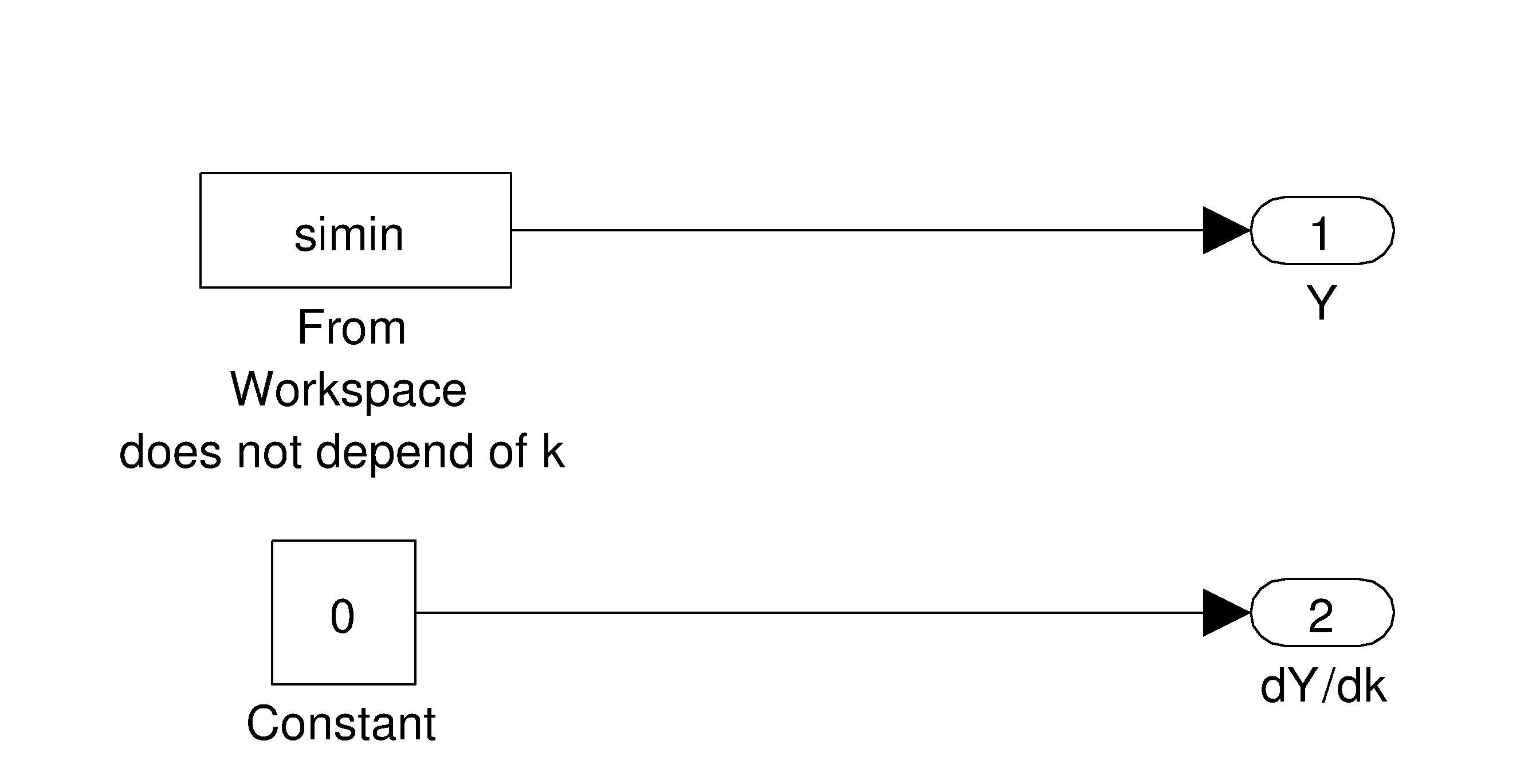}
\caption{J2 rule  derivative block independent of the parameter}
\label{J2_rule}
\end{center}
\end{figure}

All blocks unaffected by the derivative flow and not depending on
derivative parameter can be considered as a constant source equal to
zero like constant block, sourcefrom workspace , etc In other words,
when the inputs do not carry the flow derivative with respect to
derivative parameter, the block does not appear in the derivative
block diagram. This rule is useful for simplifying the derivative
model. (see \ref{J2_rule})

\subsubsection{Block rules M1, M2, M3, M4, M5}

\paragraph{M1}
 All the inputs/outputs of each
of block and the sub-block of the original scheme should be accessible
at every step time of the simulation. Because it is necessary to have
the mathematical function of each block to obtain the derivative
scheme. Moreover, each subsystem is increased of the derivative
flow. For instance, the subsystem with 2 inputs and 1 output
differentiated with respect of 2 parameters k2 and k2 gives the
following:

\begin{figure}[!h]
\begin{center}
\includegraphics[scale=1]{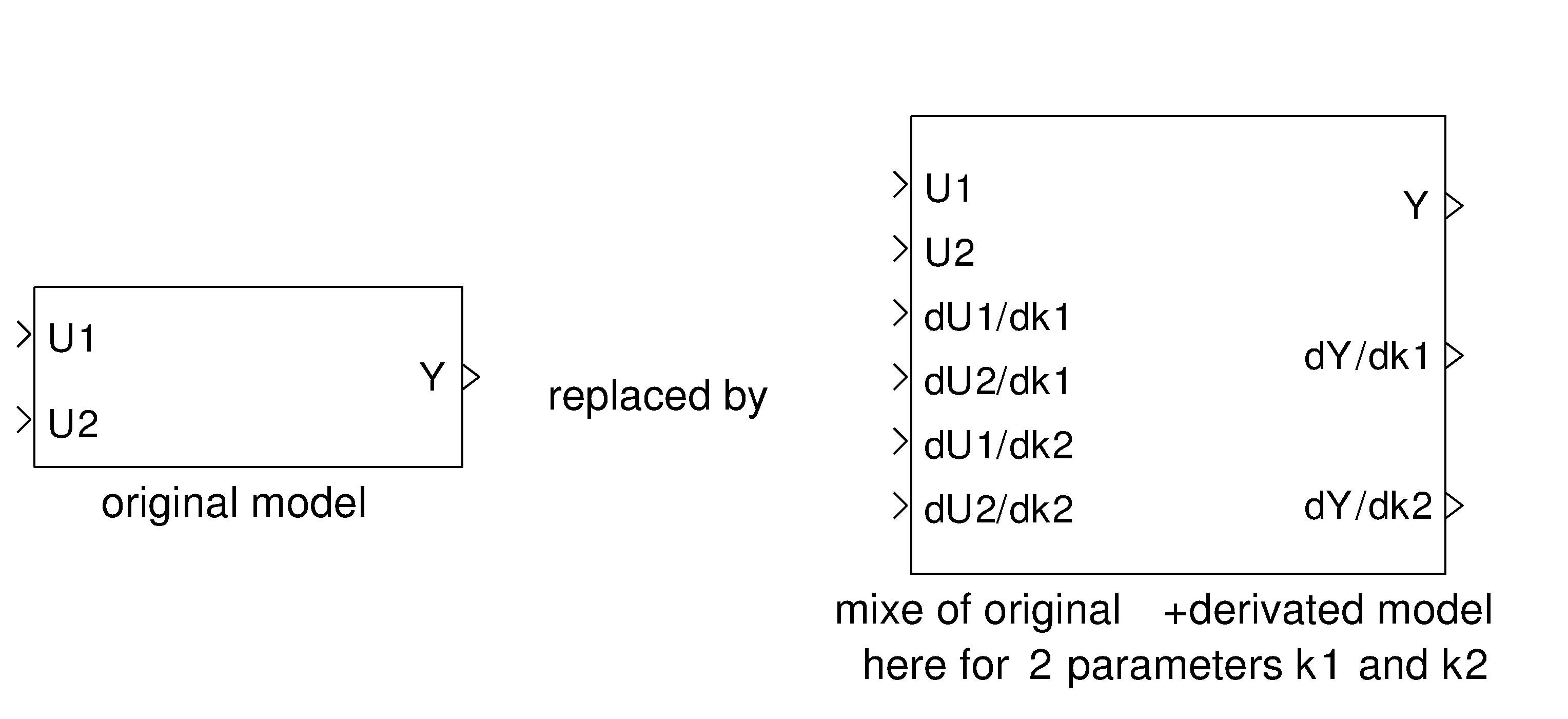}
\caption{Derivative subsystem}
\label{M1_rule}
\end{center}
\end{figure}

\begin{figure}[!h]
\begin{center}
\includegraphics[scale=.7]{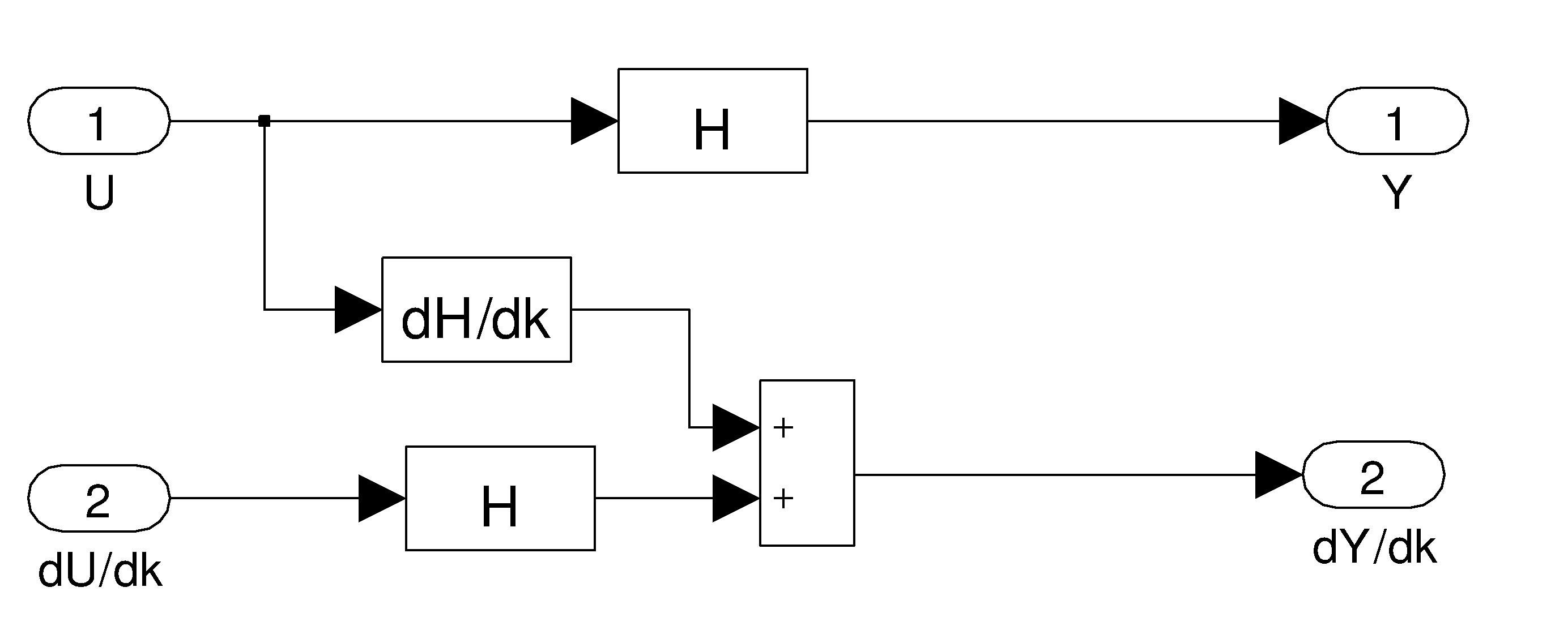}
\caption{M2 rule for alinear block  depending on of the parameter $k$}
\label{M2_rule}
\end{center}
\end{figure}

\paragraph{M2}
 For all linear blocks H(k) in U through by differentiated flow we
can built the following scheme that contains the original block
diagram increased of derivative with respect to a parameter k.(Fig.~\ref{M2_rule})

\begin{figure}[!h]
\begin{center}
\includegraphics[scale=.7]{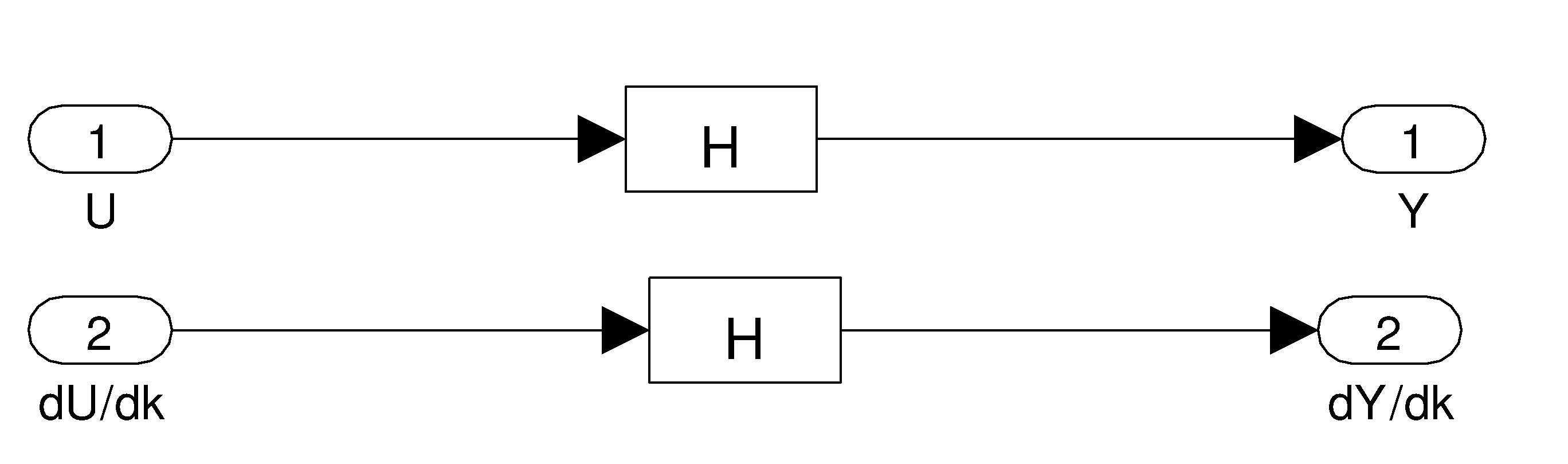}
\caption{M2 rule for linear block without parameter}
\label{linear_rule}
\end{center}
\end{figure}

If the block does not depend of derivative parameter, we obtain the
diagram below (figure \ref{linear_rule}). 
 
In fact, when the blocks are linear and do not depend on the
differentiation parameter, we just need to duplicate the model into the
derivative flow. We will increase the original model as many times as
there are parameters with respect to which differentiation is performed.

\paragraph{M3}
  For a nonlinear block we write 
 
 \begin{figure}[!h]
\begin{center}
\includegraphics[scale=0.8]{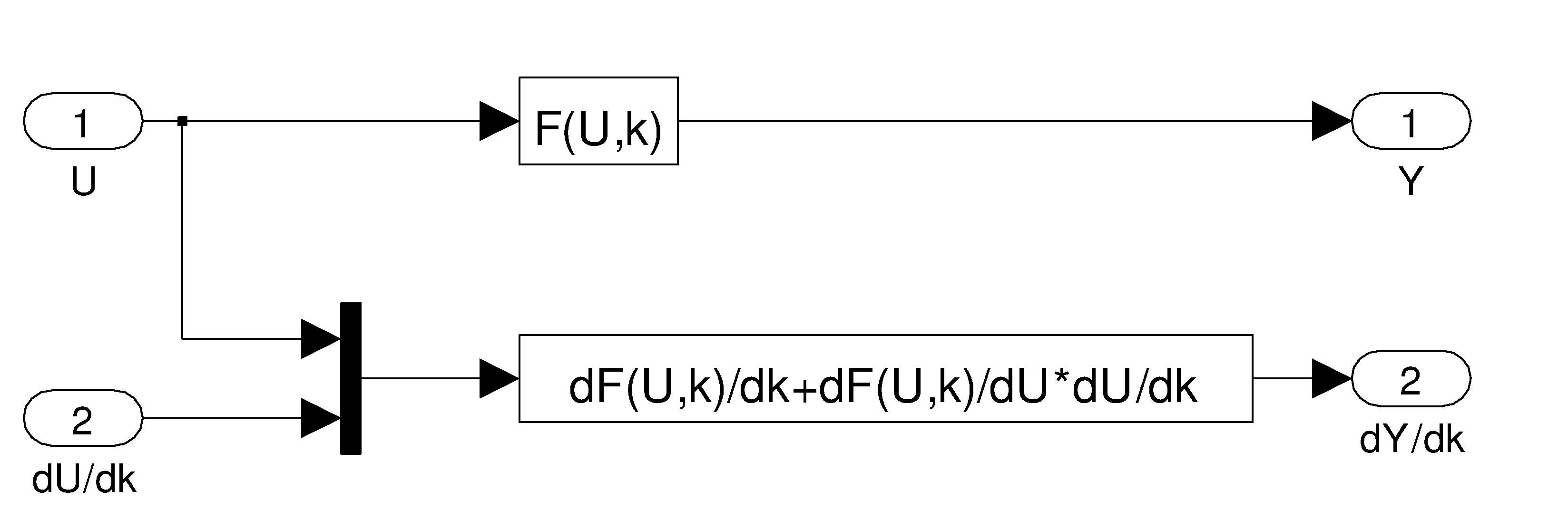}
\caption{M3 for non linear block in Laplace or Z}
\label{M3_rule}
\end{center}
\end{figure}

\begin{figure}[!h]
\begin{center}
\includegraphics[scale=0.7]{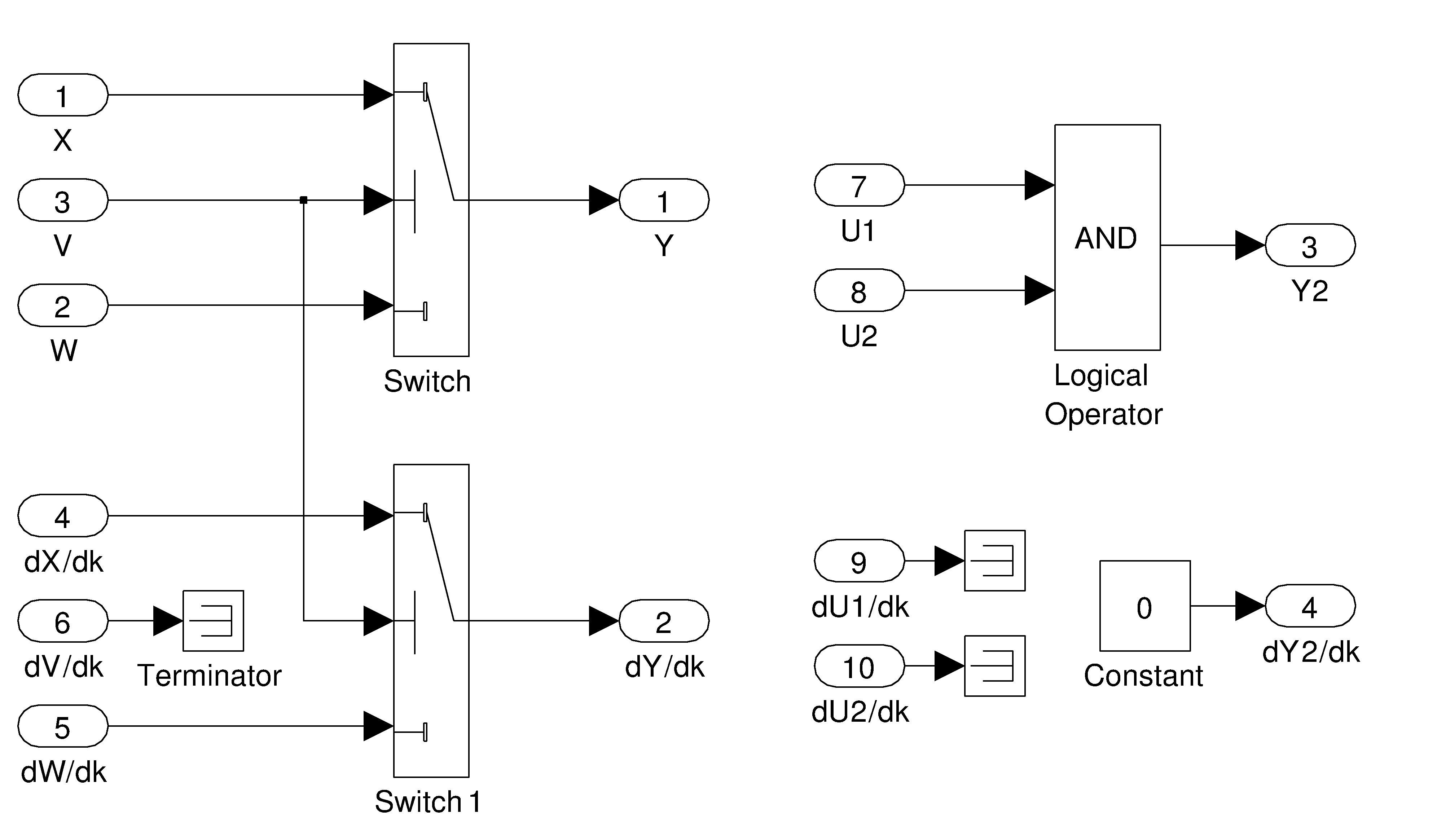}
\caption{M4 rule on switch and AND block}
\label{M4_rule}
\end{center}
\end{figure}

 In this case, it may be necessary to use a computer algebra system to
 compute the derivatives, accordiing to the mathematical definition of
 the block.
\paragraph{M4} 
 For a conditional block (Switch, hysteresis, max, min, trigger
subsystem, logic( and, or,etc) , stateflow..., saturated integrator
...)  which is defined by a piecewise or event function, we duplicate the
block and we keep the same logical tests as in the original block but
the outputs contain the derivative flow. According to automatic
differentiation, the threshold of logical blocks has not to be
differentiated.(figure \ref{M4_rule})

\paragraph{M5}

In the case of ``black box'' block, such as a lookup table 1D, $n$D or
\texttt{S-function}\footnote{Simulink block written outside of the
  block environment, in a computer language such as Matlab, C, C$^{++}$
  or Fortran.}, the mathematical equations are not accessible and
nothing can be done in the case of lookup tables. So we need to
retreat to finite difference. \textit{E.g.} in the case of a function
$y(x,k)=f(x,k,u(x,k))$, we obtain an evaluation of the derivative
$\frac{df}{du}=\frac{\partial f}{\partial k}+\frac{\partial
  f}{\partial u}\frac{\partial u}{\partial k}$:

\begin{align}\label{finite-diff}
\frac{dy(k,u(k))}{dk}&=\frac{f(x, k+\epsilon_{k},u(x,k+\epsilon_{k})-f(x,k,u(x,k))}
{\epsilon_{k}}\nonumber\\
&+
\frac{f(x,k,u(x,k)+\eta_{u(k)})-f(x,k,u(x,k))}
{\eta_{u(k)}}\frac{\partial u(k)}{\partial k}.
\end{align}

The best choice of the increment $\eta_{u(k)}$ of the
input $u(k)$ is the step between 2 values of $u(k)$ input values of
a lookup table. In general, the choice of the increment $\epsilon(k)$
may require many tests befor finding a proper value and it is better
to adapt the choice for each
lookup table or black box block rather than using a single value
$\epsilon$ for the whole
model.

\begin{figure}[!h]
\begin{center}
\includegraphics[scale=0.7]{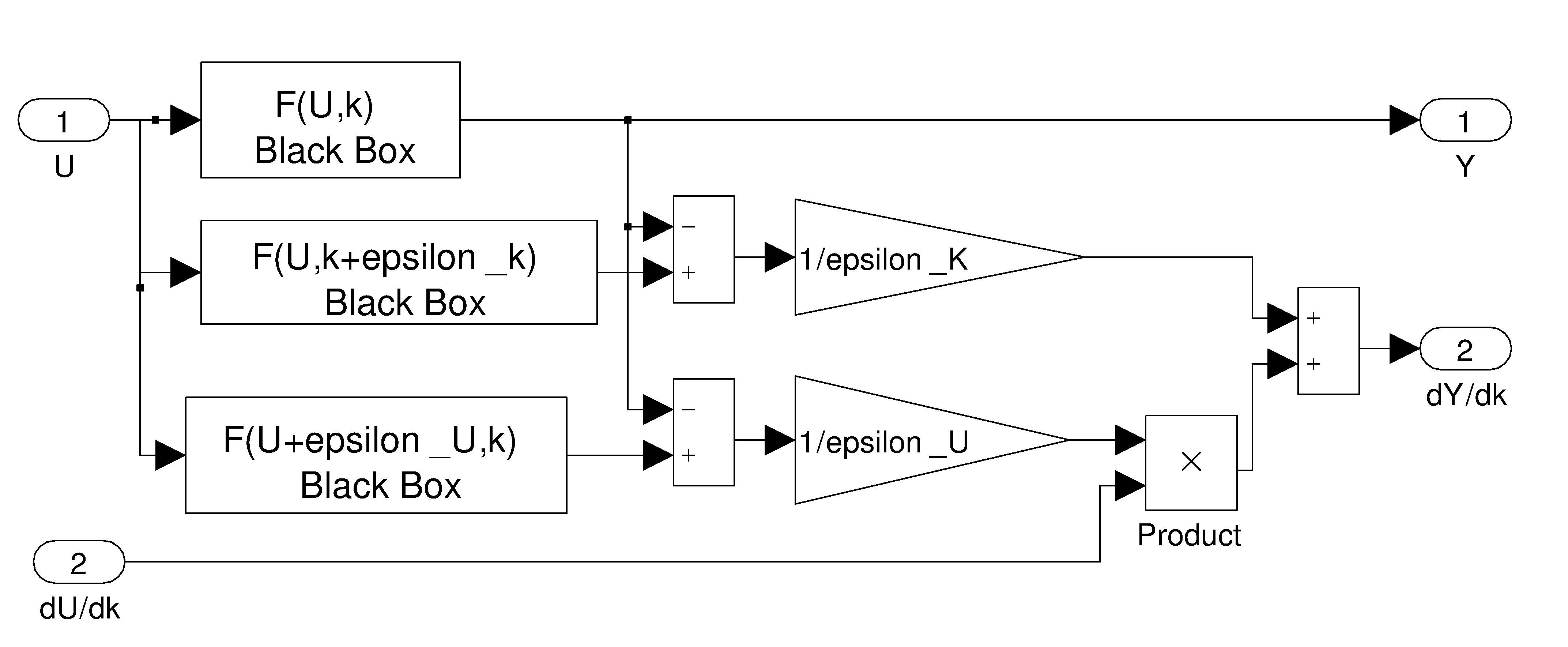}
\caption{M5 rule for Black box}
\label{M5_rule}
\end{center}
\end{figure}

All the rules introduced above are sufficient to compute the derivative model
of all Simulink models and for specifying an automata computing
the derivative like Diffedge.

\subsubsection{Other mathematical representation} 

This notion of graphic derivative may be generalized to all
mathematical objects like continuous transfer and state space but also
to discrete transfer function and state space. It is thus possible to
build the gradient of hybrid models.  For instance, in the case of
continuous state space JM's rules give the original state-space
increased of the derivative flow

\paragraph{Continuous case}

{\footnotesize
\begin{equation}\label{Space}
\left\{\begin{array}{lcl}
\dot X &= &AX + BU\\
Y &= &CX + DU
\end{array}\right. \Longrightarrow \left\{
\begin{array}{lcl}
  \left(\begin{array}{c}
  \dot X\\
  \frac{d\dot X}{dk}
  \end{array}\right)&=
  &\left[
  \begin{array}{cc}
    A &0\\
    \frac{dA}{dk} &A
  \end{array}
  \right]
  \left(\begin{array}{c}
  X\\
  \frac{dX}{dk}
\end{array}\right)+
\left[
\begin{array}{cc}
B &0\\
\frac{dB}{dk} &B
\end{array}
\right]
\left(\begin{array}{c}
U\\
\frac{dU}{dk}
\end{array}\right)\\
  \left(\begin{array}{c}
  Y\\
  \frac{dY}{dk}
  \end{array}\right)&=
&\left[
\begin{array}{cc}
C &0\\
\frac{dC}{dk} &C
\end{array}
\right]
\left(\begin{array}{c}
X\\
\frac{dX}{dk}
\end{array}\right)+
\left[
\begin{array}{cc}
D &0\\
\frac{dD}{dk} &D
\end{array}
\right]
\left(\begin{array}{c}
U\\
\frac{dU}{dk}
\end{array}\right)
\end{array}
\right.
\end{equation}
}

\paragraph{Discrete case} 

For the Discrete case, we have the same structure with
the following notation: $\dot X=X_{n+1}$ and $X=X_{n}$ and idem for
$U$ and $Y$. 

In Simulink The Discrete Transfer function
block implements the z-transform transfer function described by
$z^{n}$ or $z^{-n}$
power with the following equations in z:

\begin{equation}\label{zT}
H(z) =
\frac{b_{0}+b_{1}z+\cdots+b_{m}z^{m}}{a_{0}+a_{1}z+\cdots+a_{m}z^{m}}\quad
\hbox{with}\quad  m\le n.
\end{equation}
 
\subsection{Other examples}

Simulink blocks have often dissimulate complex structures. It may then
be necessary to rewrite theses blocks using elementary blocks to be
able to apply the rules and compute the derivative of blocks such as
\texttt{Min}, \texttt{Max}, etc.

\paragraph{Discontinuous block:}
For example, we propose to differentiate the
Saturation Dynamic block (the bounds range of the input signal to
upper and lower saturation values). The input signal outside of these
limits saturates to one of the bounds low or up port. We suppose, the
bounds do not depend of the parameter.
 
\begin{figure}[!h]
\begin{center}
\hbox
to\hsize{\hss\includegraphics[scale=1]{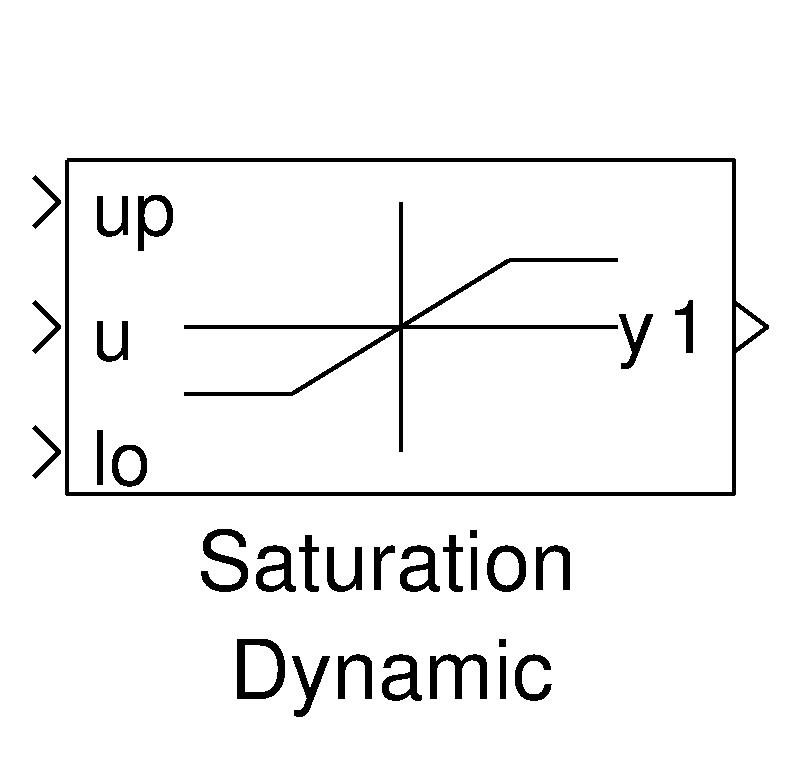}
\hss\hss\includegraphics[scale=1]{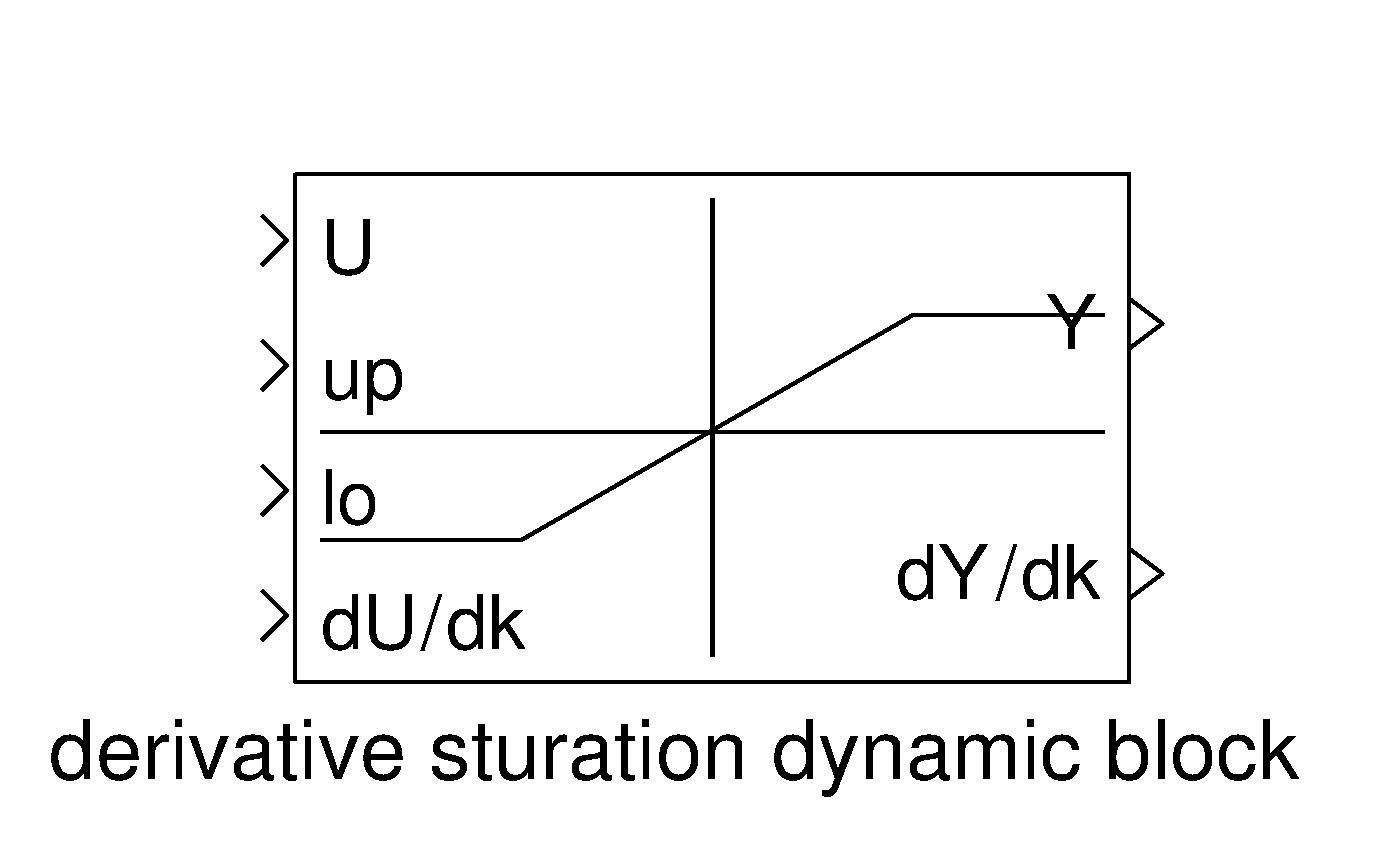}\hss}
\caption{Saturation Dynamic block (left) and its derivative (right)}
\label{saturation}
\end{center}
\end{figure}

\begin{figure}[!h]
\begin{center}
\includegraphics[scale=0.5]{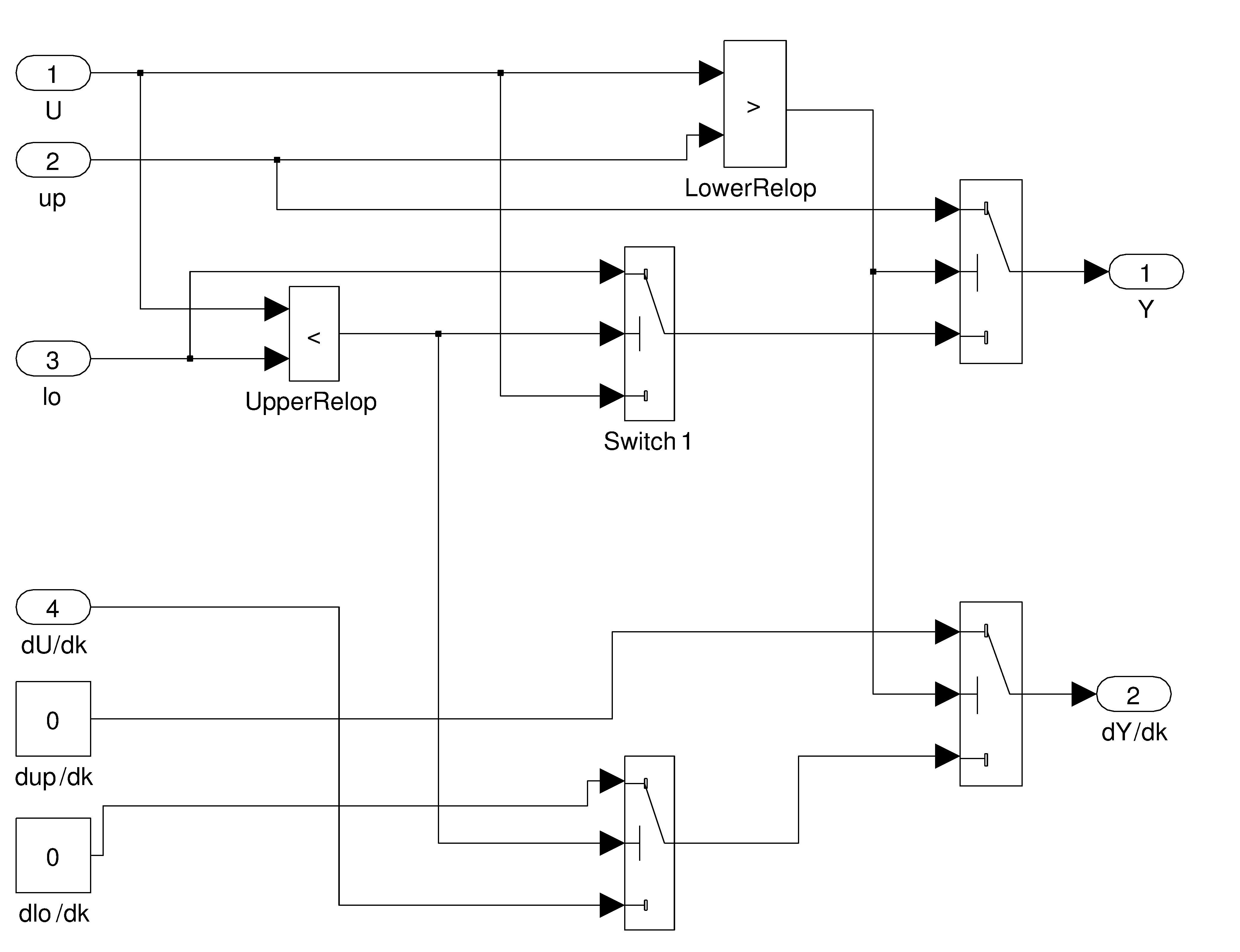}
\caption{Inside the Derivative of Saturation dynamic  block}
\label{saturatio_der}
\end{center}
\end{figure}

 \paragraph{Discrete block.} 

\begin{figure}[!h]
\begin{center}
\includegraphics[scale=0.7]{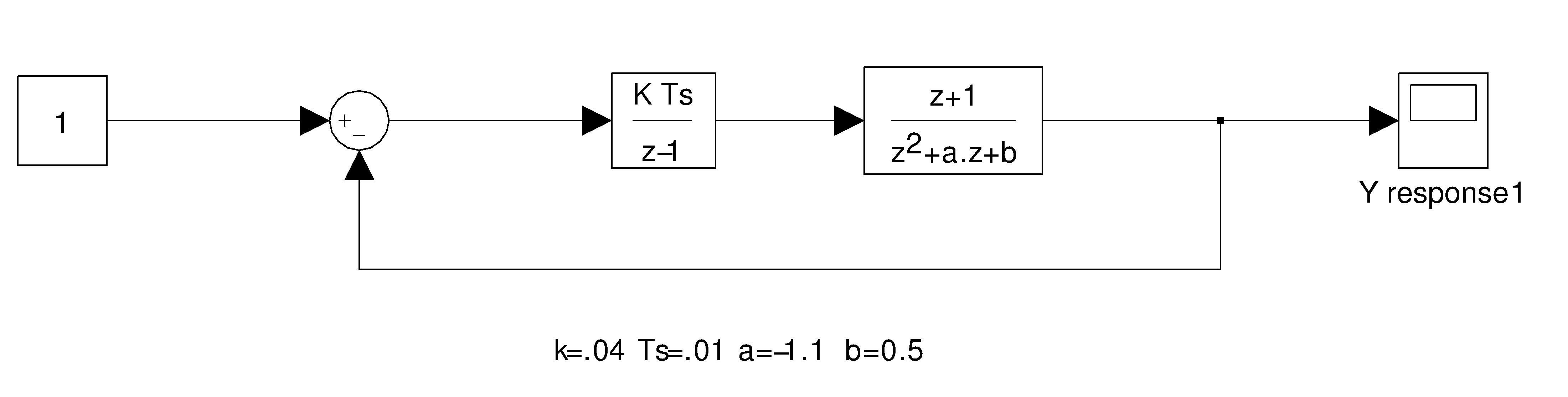}
\caption{Discrete representation}
\label{discrete}
\end{center}
\end{figure}

We want to study the sensibility of the parameter k inside of discrete
integrator, it is necessary to compute the derivative (Fig.~\ref{discrete})

\begin{figure}[!h]\label{d-discrete}
\begin{center}
\includegraphics[scale=0.7]{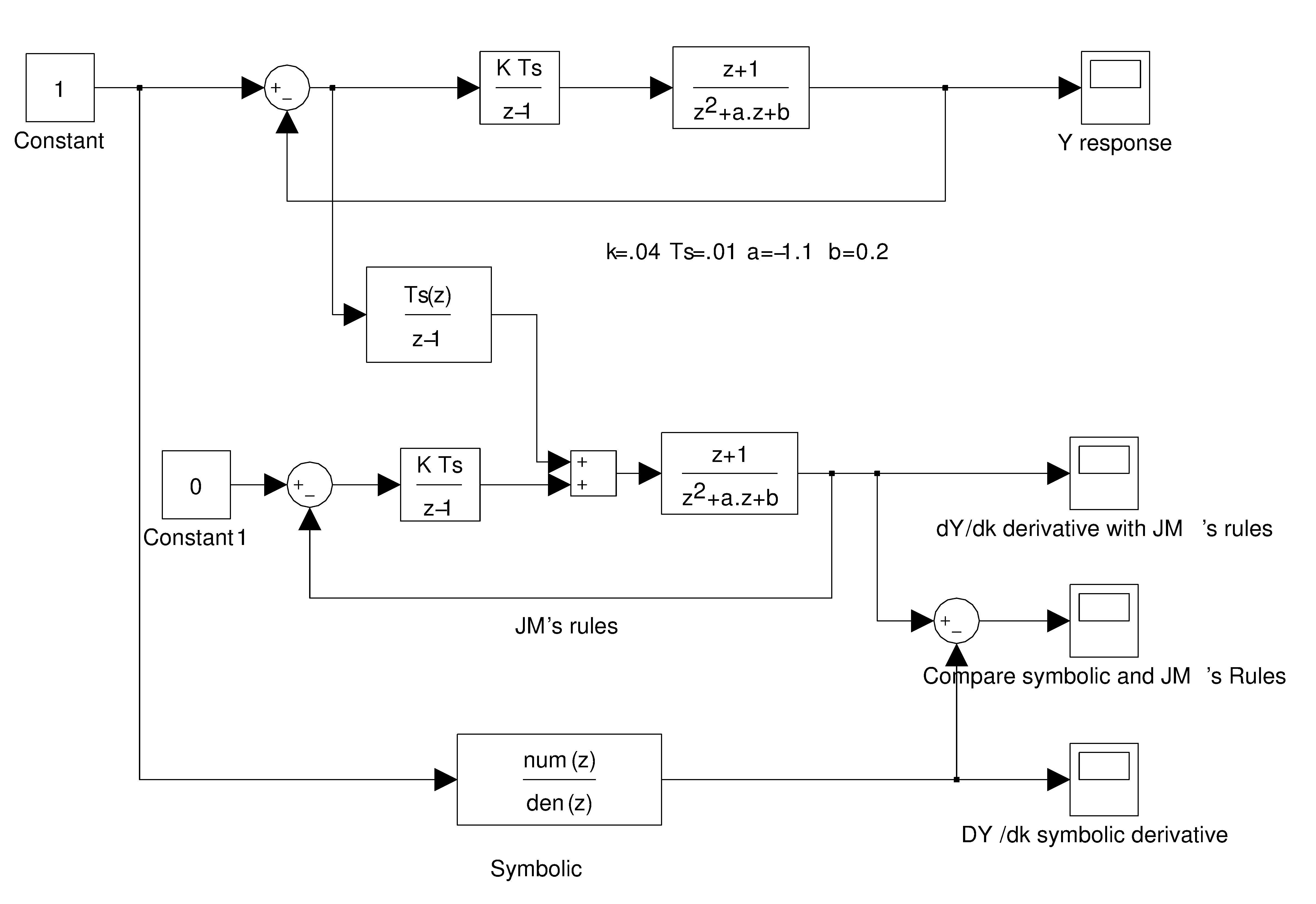}
\caption{Discrete derivative representation}
\end{center}
\end{figure}

Application of rules gives the Fig.~\ref{d-discrete} 
 For computing the derivative of the (Fig.~\ref{discrete}),  we duplicate the original
 model. We just put the source at zero and link the derivative of
 the integrator with respect to K just after the add block then
 finally we put a additional block in the derivative flow. 

And at last, we compare the ADGM with the evaluated derivative. This
last one step is very complicated to obtain. We have to evaluate the
close loop, to compute the derivative with respect to
k and to convert it into a rational fraction in canonic form that will
be entered in the dialogue discrete box of Simulink. The next formula
expresses the formal derivative and is implemented in the ``Symbolic''
block of fig.~\ref{d-discrete}.

\begin{equation}\label{eq_d_discret}
\frac{dY(z)}{dk}=
\frac{Ts\left(z^{4}+az^{3}+\left(b-1)\right)z^{2}-az+KTs-b\right)-KTs^{2}}
{\left(z^{3}+\left(a-1\right)z^{2}+\left(KTs-a+b\right)z+KTs-b\right)^{2}}.
\end{equation}
 
The result obtained with ADGM is exactly equal to what we get frrom the integration of equation~\ref{eq_d_discret},
with less effort.


\begin{figure}[!h]
\begin{center}
\includegraphics[scale=0.75]{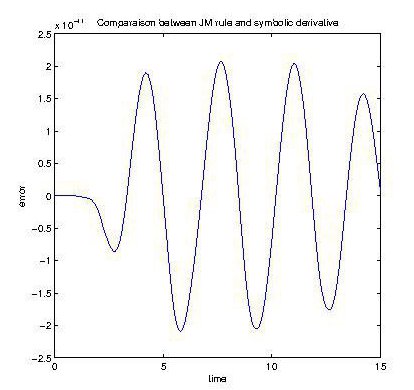}
\caption{Error between the true and ADGM derivative	 }
\label{discrete_errorr}
\end{center}
\end{figure}

 The best way for computing the derivative is to use JM's rules It is
 very simple, efficient, fast and useful. This method is also a good
 way for optimizing Finite Response Input (FIR) filters\footnote{Such
   filters are known to have usefull property such as inherent
   stability.} with saturation.

\subsection{An example of optimization}

\begin{figure}[!h]
\begin{center}
\includegraphics[scale=0.7]{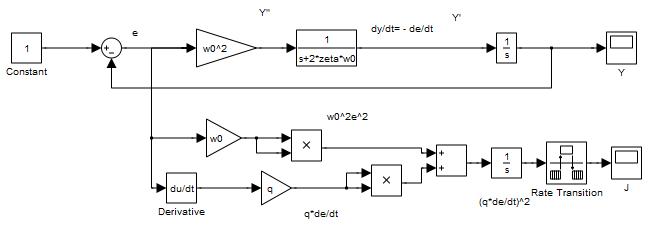}
\caption{Second order with cost function J }
\label{secondorder}
\end{center}
\end{figure}

We conclude this section with a classical example. We consider a
second order system and try to minimize the energetic cost of the
transitory behaviour during a step action. Our second order system
is the following (Fig.~\ref{secondorder}).
\begin{eqnarray}\label{ordre2}
x''&=& - 2\zeta\omega x' - \omega^{2}x + \omega^{2}u(t)\\
y  &=& x
\end{eqnarray}
The initial conditions are $x(0)=0$ and $x'(0)=0$ and $u(t)$ the
Heaviside function $u(t)=0$ if $t<0$ and $u(t)=1$ if $t\ge 0$.

We want to minimize 
$$
J(\zeta)=\int_{0}^{\infty}( \omega^{2}e(t)^{2} + q^{2}e'(t)^{2} ) \td t,
$$
with $e(t)=x(t)-u(t)$.


In this simple case, a simple Maple computation will provides the
optimal value of $\zeta$  with $q=1$ and $\omega=1$ is  $\zeta=\frac{sqrt(2)}{2}= 0.7071067811865475 $ with 16 Digits.
We will use it to illustrate how Diffedge can be
used for optimization. For this, we need to compute 
$(\partial J/\partial \zeta)(\zeta)$, in order to use the
Levnberg--Marquardt algorithm, implemented in Matlab fsolve
function. The time of simulation is  10~sec.

Our goal is to compare AD and finite difference in real world
condition(by example embedded optimization on micro computer), many practical examples encountered in engeneering practice
being close to this archetypic problem wherever the sample time of the
observation is large. 

\begin{figure}[!h]
\begin{center}
\hbox to\hsize{\hss\includegraphics[scale=0.4]{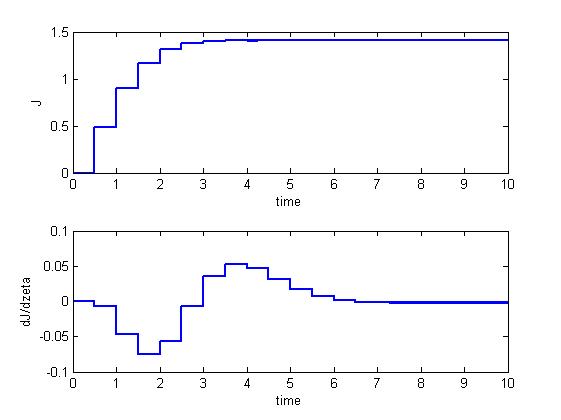}
\hss\hss\includegraphics[scale=0.4]{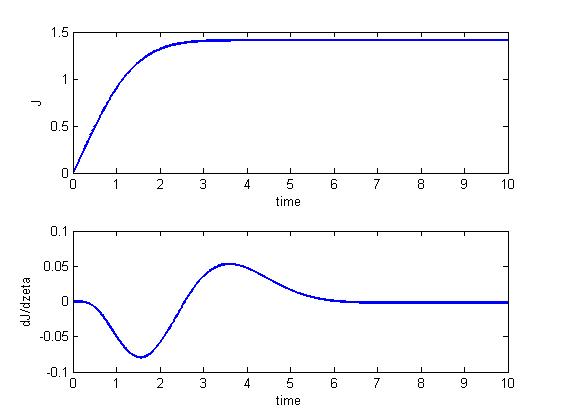}\hss}
\caption{Tsample=$0.5$ seconds (left) and Tsample$=0.005$ seconds (right)}
\label{response1}
\end{center}
\end{figure}

\paragraph{Remark.} One must be carefull to the
sensibility of the computation of the integral $J$ with respect to the
time observation frequency: not enought points will give a poor
precision in optimization, but too many points can cumulate
inacuracies and finite difference method will be very sensitive to
such computational noise.  The time observation is given by the block
\texttt{rate transition} with the variable $Tsample$. Here after
(fig.~\ref{response1}, we show two responses
with differente values of $Tsample$. One may notice that there are too
few points in the sample here to use finite difference but that AGDM
still works.

\paragraph{Differentiation.}
The corresponding block diagram of these equations can be written in the following way (Fig.~\ref{secondorder}).
Following classical engeneering practice, one may use Heaviside
operational calculus and represent our equations by a unit feedback of the
open loop system described by the transfer function
$$
\frac{\omega^{2}}{s(2\omega\zeta+s)}.
$$

\begin{figure}[!h]
\begin{center}
\includegraphics[scale=0.8]{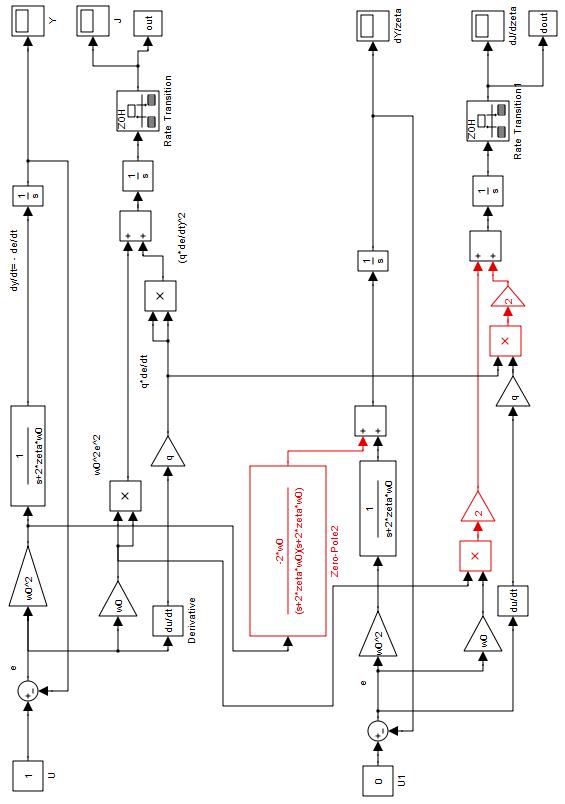}
\caption{Second order derivative with respect to $\zeta$ }
\label{dsecondorder}
\end{center}
\end{figure}

In this model, the cost is simulated together with the initial
model. One may notice that, for the sake of genericity, the derivative
$e'(t)$ is computed using the block \texttt{derivative}, that uses
finite difference, that may penalize accuracy. But, if $u$ were not
constant for $t\ge 0$, we would have no other choice, as $e'=x'+u'$,
except if $u$ is itself defined in such a way that AD can be used for
it.
The differentiated model, described by the following
(fig.~\ref{secondorder}). is obtained by duplicating the initial model
(fig.~\ref{dsecondorder}).  and applying the rules:

--- J2 on the constant;

--- M2 on the linear block depending on $\zeta$;

--- M3 on the nonlinear blocks like multiplication.

Modifications with respects to the initial model appear in red.

The optimization program to be used here is the following. The output
$\td J/\td \zeta$ of our differentiated block diagram is embedded in the
\texttt{myfcost} function. 

\begin{verbatim} 
function [J,Jacobian] = myfcost(zeta)
zeta=min(zeta,3);zeta=max(.001,zeta);
assignin('base','zeta',zeta)
sim('mymod');
out=eval('base','out');   % get J in the  workspace 
dout=eval('base','dout'); % get dJ/dzeta  workspace 
J=out;                    % Objective function 
if nargout > 1            % Two outputs argument
Jacobian = dout;          % Jacobian of the function (ADGM) 
end

\end{verbatim}

To launch optimization, where $\zeta$ is initialized to the value
$0.1$, one uses the following syntax with option \texttt{on} for
ADGM/Symbolic or \texttt{off} for finite differences, when computing
the Jacobian. 

{\small
\begin{verbatim}
[zeta_opt,fval,exitflag,output]=fsolve(@myfcost,[.1],optimset('Jacobian','on'))
\end{verbatim}
}

One may compare the obtained value to the theoretical
solution of  $\zeta$.

{\scriptsize
\noindent{\tt
\begin{tabular}{|l|l|l|}\hline
Rate transition    & Matlab jacobien       & ADGM jacabien  \\
(time observation) & (finite difference)   & (symbolic)	  \\
                   &                       &  Theorical $\zeta= 0.7071067811865475 $\\
\hline
0.5                & Zeta\_opt= 0.1        & Zeta\_opt=.70698585\\
                   & 1 iterations          & 9 iterations\\
                   & funcCount: 22	   & funcCount: 23	 \\
\hline
0.1                & Zeta\_opt= 0.10154269 & Zeta\_opt=0.71180603\\
                   & 8 iterations          & 8 iterations\\
                   & funcCount: 40	   & funcCount: 25	 \\
\hline
0.07               & Zeta\_opt= 0.10089164 & Zeta\_opt= 0.70857782\\
                   & 4 iterations          & 7 iterations\\
                   & funcCount: 29	   & funcCount: 24\\
\hline	
0.04	           & Zeta\_opt= 0.10062591 & Zeta\_opt= 0.71130072\\
                   & 5 iterations          & 9 iterations\\
                   & funcCount: 32	   & funcCount: 22\\
\hline	
0.01	           & Zeta\_opt= 0.70625955 & Zeta\_opt= 0.70671083\\
                   & 9 iterations          & 9 iterations\\
                   & funcCount: 29	   & funcCount: 19\\
\hline
0.005              & Zeta\_opt= 0.70630892 & Zeta\_opt= 0.70628714\\
                   & 9 iterations          & 9 iterations\\
                   & funcCount: 29	   & funcCount: 29\\
\hline
\end{tabular} 
}}

The number of observation points does not improve the result. In our
case the solver is ode45 and the relative tolerance is
$rtol=10^{-7}$. This means that the computed state ($J$) is accurate
to within 0.00001\% It is not possibe to get better precision than
around $\approx 10^{-4}$ even using symbolic jacobian. But
we notice that the optimization with a symbolic
jacobian is more reliable.
  
\section{Testing the precision}

It seems safe to complete this presentation with a few hint to test
the robustness of the obtained results. Most of the time, it is
of course impossible to compare results using an extra tool with
certified reliability.  And if
the result of the other tool is not somehow ``certified'', a third one
would be required, in case of
discrepancies, assuming that if two methods agree they are likely to
be right.

However, any numerical tool may be tested on a set of known systems
that already may give an idea of its possibilities and help to test it
during its development.

\subsection{Symbolic computation}

The possibilities are obvious enough to dispense us of lengthy
developments, but as most differential systems do not have closed form
solution, let us stress that the safest way would be a start by the
solution and reconstruct the system. Chained systems provide an easy
way to build systems of increasing size and complexity when keeping
symbolic resolution fast. 

\subsection{Flat systems}

Flat system, already menstionned above (see \ref{Flat}), are a special
class of differential systems with a closed form solution, including
well known chained examples of arbitrary size, like the car with
$n$-trailer \cite{fliess1993} or many discretization of flat inifinite
dimensional systems \cite{Ollivier2001}. Such examples where used with
success during the tests of the rocket engine simulator Carins
\cite{Ordonneau2006}, allowing to detect a bug in Maxima.

\subsection{Interval arithmetics}

As interval arithmetic can garanty that the result belongs to the
result interval, it may also be used to test any numerical tool. The
computer algebra system \href{http://www.mathemagix.org/www/mmdoc/doc/html/main/index.en.html}{Mathemagix} includes a package \href{http://www.mathemagix.org/www/numerix/doc/html/index.en.html}{\texttt{numerix}} for
intervals and also balls with certified arithmetics
\cite{HoevenLecerf2016,Hoeven2016}.

\section*{Conclusion}

We have shown that, if the advantages of AD are obvious, one needs to be
carefull in many cases of great pratical interest, where its naive use
may lead to inaccuracies. A clever use requires a high level knowledge
of what subroutines are made for, that is not just their syntax but
also semantics. \textit{E.g.} we have shown that differentiating a
loop heavily depends on the fact that the loop is assumed to converge
to a fixed point, or not. In the worse case, automatic procedures will
be unable to recover the semantic hidden behind the syntax, which may
lead to unavoidable inaccuracies. The best solution would be to
generate numeric code from symbolic formulae, allowing more
flexibility if new expressions, such as a derivatives, are
required. But the available tools are still limited.

Simulink block diagrams offer a perfect illustration of high level
object that may---and must---be differentiated as high level object,
leading to better accuracy and keeping the benefit of their wide range
of possibilities, including translation in Fortran or C to incorporate
of real time optimization routines in embeded code. To some extend,
Simulink block diagram description offers a better and more flexible
framework to compute gradients with AGDM methodology that what is
offered, not only by low level coputer languages like C or Fortran,
but also by computer algebra systems like Maple, with the restriction
of discontinuities that cannot be handled properly and require the use
of continuous approximations using $\arctan$.

Being able to produce numerical programs that could be
efficient and flexible, \textit{i.e.} that could easily produce extra
outputs such as derivatives, remains a challenge for software
engineering. Producing numerical code from computer algebra
specification is a part of the answer, but we need to adapt our tools
to produce not only matematical formulas, but programs that must be
fast and well conditionned. For the present time, we still have to
cope with the necessity of pluging AD features on softwares that have
not been thought in advance to facilitate the task or even make it
possible and safe.

\end{document}